\documentclass[english,twocolumn]{revtex4}
\usepackage[T1]{fontenc}
\usepackage[latin9]{inputenc}
\usepackage{graphicx}
\usepackage{amssymb}

\makeatletter

\providecommand{\tabularnewline}{\\}

\makeatletter

\makeatletter

\usepackage{float}

\makeatletter

\usepackage{float}

\makeatletter

\usepackage{float}

\makeatletter

\usepackage{float}

\makeatletter

\usepackage{float}

\makeatletter

\usepackage{float}

\makeatletter

\usepackage{float}

\makeatletter

\usepackage{float}

\makeatletter



\makeatother

\makeatother

\makeatother

\makeatother

\makeatother

\makeatother

\makeatother

\makeatother

\makeatother

\makeatother

\makeatother

\usepackage{babel}

\begin{document}

\title{Counterflow and paired superfluidity in one-dimensional Bose mixtures
in optical lattices}

\author{Anzi Hu$^{1}$, L. Mathey$^{1}$, Ippei Danshita$^{2,3}$, Eite Tiesinga$^{1}$,
Carl J. Williams$^{1}$ and Charles W. Clark$^{1}$}

\affiliation{$^{1}$Joint Quantum Institute,University of Maryland and National
Institute of Standards and Technology,Gaithersburg, MD 20899\\
 $^{2}$Department of Physics, Faculty of Science, Tokyo University
of Science, Shinjuku-ku, Tokyo 162-8601, Japan\\
 $^{3}$Department of Physics, Boston University, Boston, MA 02215\\
 }

\begin{abstract}
We study the quantum phases of mixtures of ultra-cold bosonic atoms
held in an optical lattice that confines motion or hopping to one
spatial dimension. The phases are found by using Tomonaga-Luttinger
liquid theory as well as the numerical method of time evolving block
decimation (TEBD). We consider a binary mixture with repulsive intra-species
interactions, and either repulsive or attractive inter-species interaction.
For a homogeneous system, we find paired- and counterflow-superfluid
phases at different filling and hopping energies. We also predict
parameter regions in which these types of superfluid order coexist
with charge density wave order. We show that the Tomonaga-Luttinger
liquid theory and TEBD qualitatively agree on the location of the
phase boundary to superfluidity. We then describe how these phases
are modified and can be detected when an additional harmonic trap
is present. In particular, we show how experimentally measurable quantities,
such as time-of-flight images and the structure factor, can be used
to distinguish the quantum phases. Finally, we suggest applying a
Feshbach ramp to detect the paired superfluid state, and a $\pi/2$
pulse followed by Bragg spectroscopy to detect the counterflow superfluid
phase. 
\end{abstract}
\maketitle

\section{Introduction}

Bose-Einstein condensation~\cite{BEC} is a fascinating many-body
phenomenon. It demonstrates the significance of quantum statistics
at low temperature. Identical bosons can occupy the same single particle
state and are in fact more likely to do so than classical particles.
At a critical temperature, a gas of bosons undergoes a phase transition
towards a state in which a macroscopic fraction of the particles occupy
the lowest energy state, creating a condensate. Such a state was realized
in ultra-cold atom systems in~\cite{BEC-exp}, demonstrating that
the technology of cooling and manipulating atoms had reached a level
of control with which novel states of matter could be generated and
studied.

In the case of a Fermi gas, the Pauli exclusion principle prevents
such a phenomenon to occur, because no single particle state can be
more than singly occupied. However, the phenomenon of condensation
can still occur in Fermi systems via a different mechanism: fermions
can form pairs to create composite bosons. The bosonic particles then
form a condensate of pairs. Conventional superconductors, for example,
were understood as a condensate of electron pairs~\cite{BCS}. In
ultra-cold atoms, fermionic condensates of this type were created
in~\cite{BCS-BEC}.

Interestingly, this mechanism of condensation of pairs is not limited
to fermionic systems but can occur in bosonic systems as well. In
fermionic systems, formation of Bosonic pairs necessarily occurs before
condensation. In bosonic systems this mechanism can be favored energetically,
and will typically be in competition with single particle condensation.

\begin{figure}
\includegraphics[width=4.3cm]{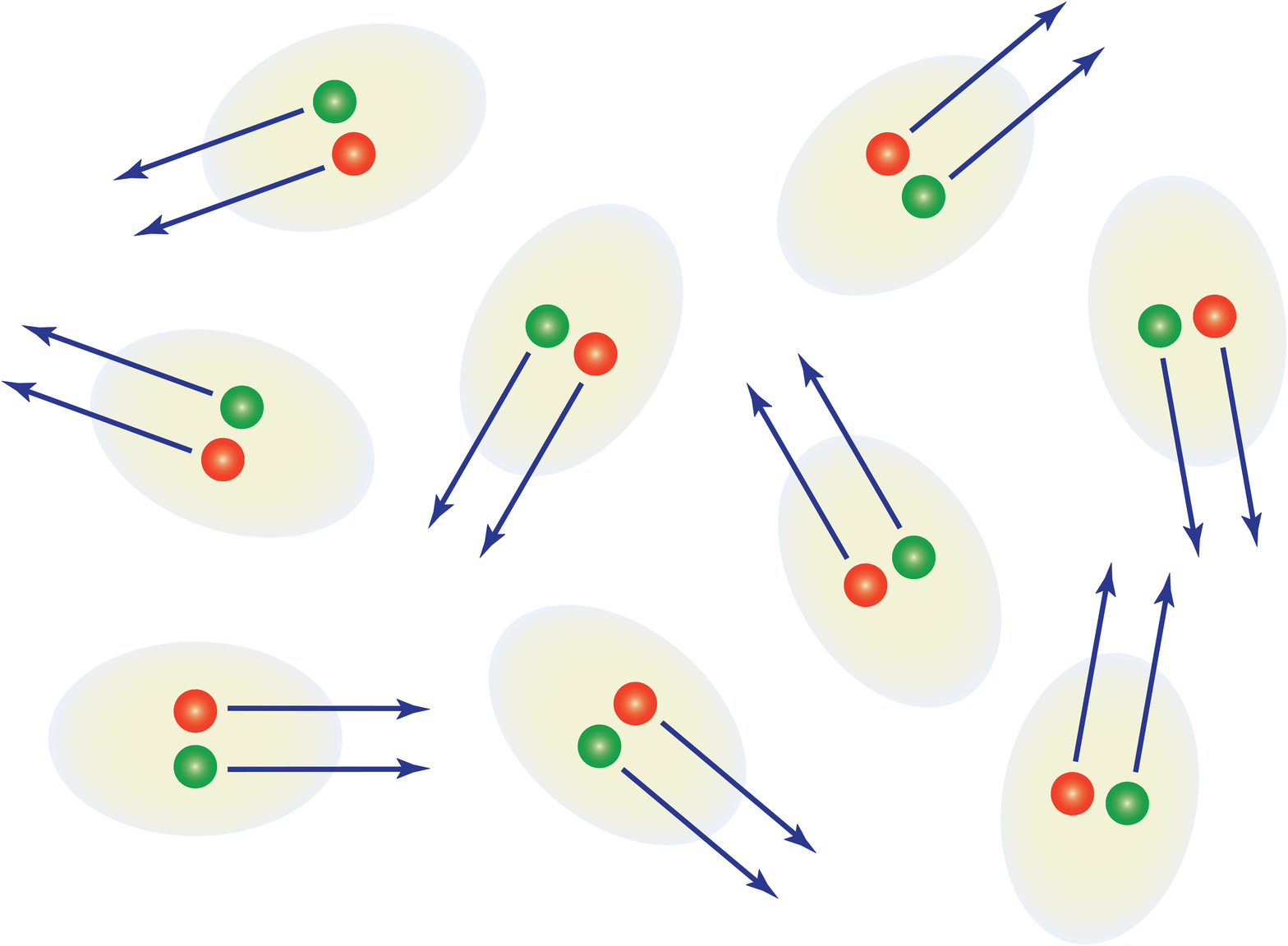}

\caption{\label{fig:Illustration-PSF} Sketch of a condensate of pairs. Atoms
of each species (red/green) pair together and form a paired superfluid
(PSF) state.}

\end{figure}

\begin{figure}
\includegraphics[width=4cm]{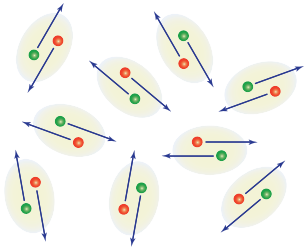}

\caption{\label{fig:Illustration-CFSF} Sketch of a condensate of anti-pairs.
Here, atoms of one species are strongly anti-correlated with atoms
of the other species, creating a counterflow superfluid (CFSF) state.
These composite bosons can also be thought of as a pair of one atom
of one species and one hole of the other species.}

\end{figure}

In~\cite{KuklovCFSF,KuklovDiagram}, two types of composite bosons
were predicted for a binary Bose mixture in a optical lattice: pairs
and anti-pairs. For attractive mutual interactions, a bosonic mixture
can form pairs of atoms which then form a paired superfluid (PSF)
state, as is visualized in Fig.~\ref{fig:Illustration-PSF}. For
repulsive interactions, at special fillings, the atoms can form anti-pairs,
which can be interpreted as pairs of one atom of one species and one
hole of the other species. These anti-pairs can then generate a counterflow
superfluid (CFSF) state, visualized in Fig.~\ref{fig:Illustration-CFSF}.
Most of their simulations were performed for two dimensional systems.

Quantum phases of atoms in optical lattices have been experimentally
studied. Following the prediction by Jaksch et al. in~\cite{Bose-Hubbard},
the Mott insulator (MI) to superfluid (SF) transition was realized
in Ref.~\cite{MI-SF(3D)} in a three dimensional lattice. In~\cite{MI-SF(1D)}
this transition was achieved in 1D. More recently, Ref.~\cite{MI-SF(2D)}
observed the two dimensional (2D) transition.

In one-dimensional gases quantum phases have quasi-long range order
(QLRO), rather than true long range order. QLRO of an operator $O(x)$
is defined as follows: The correlation function $R(x)=\langle O^{\dagger}(x)O(0)\rangle$
falls off algebraically as $R(x)\sim|x|^{\alpha-2}$ as $|x|\to\infty$
with $\alpha>0$. Various order parameters $O(x)$ will be defined
in the text. In contrast in higher dimensional bosonic systems correlation
functions can have true long range order, where correlation functions
approach a finite value. Power-law scaling in a 1D optical lattice
has been observed in~\cite{BlochTonks}. They observed the Tonks-Girardeau
regime of strongly interacting bosons.


%

In this paper we consider a two-component Bose mixture held in an
optical lattice that only allows atoms to hop in one spatial dimension.
We ask the question of how the superfluid as well as other phases
or orders can be realized. We assume that the two species of the mixture
have the same filling $\nu$, restricted to the range $0<\nu\leq1$.
The phase diagram of these mixtures is determined using Tomonaga-Luttinger
liquid theory~\cite{giamarchi_book}, which gives the universal
phase diagram in terms of a few effective parameters. Based on the
univerisal phase diagram, we generate the numerical phase diagram
using the time-evolving block decimation (TEBD) method~\cite{Vidal,daley,danshita1,danshita2}.
With these two approaches we find that CFSF can exist for $\nu=1/2$
(half-filling) and repulsive interaction, whereas PSF can exist for
$\nu<1$ and attractive interaction (see also \cite{Santos}).

We also find that charge density wave (CDW) quasi-order can coexist
with both PSF and CFSF, as well as single particle superfluidity (SF).
The regimes in which CDW and SF quasi-order coexist constitute a quasi-supersolid
phase~\cite{supersolid_LM,supersolid}. Similarly, the regimes where
CDW and PSF quasi-order coexist is a quasi-supersolid of pairs and
in the case of CFSF, a quasi-supersolid of anti-pairs. Previous work
has predicted coexistence of CDW and PSF for 1D Bose mixtures~\cite{Commensurate,supersolid_LM}
and bilayer 2D lattice bosons with long-range interactions~\cite{Pair-SS-Lewenstein},
and that of CDW and CFSF for 1D Bose-Fermi mixtures~\cite{supersolid,Mori}.

We then address the question whether PSF and CFSF can be realized
and detected in experiment. To simulate the effect of a global trap,
we numerically study a mixture confined by a harmonic trap and find
that PSF and CFSF can indeed exist in such trapped systems. Their
existence can be detected through various measurements. The PSF phase
can be detected by using a Feshbach ramp, similar to what has been
used in BEC-BCS experiments~\cite{BCS-BEC}, which generates a quasi-condensate
signal in the resulting molecules. The CFSF phase can be detected
by applying a $\pi/2$ pulse followed by Bragg spectroscopy. This
generates a quasi-condensate signal in the structure factor. Time-of-flight
expansion can also be used to show the absence of single particle
superfluidity in PSF and CFSF. Measuring the structure factor via
Bragg spectroscopy can be one way of detecting CDW order.

This paper is organized as follows: in Section \ref{sec:The-theory},
we introduce the model that is used to describe the system; in Section
\ref{sec:Luttinger-liquid-approach}, we use Tomonaga-Luttinger liquid
theory to derive the phase diagram. The numerical approach and results
are discussed in Section \ref{sec:Numerical-Approach}. Specifically,
phase diagrams of the homogeneous system are presented in Section
\ref{sec:The-phase-diagrams}, and the realization and detection of
PSF and CFSF are discussed in Section \ref{sec:Experimental-realization-and}.
We conclude in Section \ref{sec:Summary}.

\section{\label{sec:The-theory}Hamiltonian}

Ultra-cold bosonic atoms in optical lattices can be well described
by Bose Hubbard models~\cite{Bose-Hubbard}. Here, we consider a
mixture of two types of atoms confined to a one-dimensional lattice
system. The Hamiltonian of such a system is given by \begin{eqnarray}
H & = & -t\sum_{a=1,2}\sum_{i=1}^{N-1}(b_{a,i}^{\dagger}b_{a,i+1}+h.c.)+U_{12}\sum_{i=1}^{N}n_{1,i}n_{2,i}\nonumber \\
 &  & +\frac{U}{2}\sum_{a=1,2}\sum_{i=1}^{N}n_{a,i}(n_{a,i}-1).\label{eq:hamiltonian}\end{eqnarray}
 We denote the different types of atoms with index $a=1,2$, and the
lattice site with index $i$. We assume that the two species have
equal particle density $\nu\leq1$, the same intra-species interaction
$U>0$ and hopping parameter $t>0$. The inter-species interaction
is given by $U_{12}$. The operators $b_{a,i}^{\dagger}$ and $b_{a,i}$
are the creation and annihilation operators for atoms of type $a$
and site $i$ and $n_{a,i}=b_{a,i}^{\dagger}b_{a,i}$ are the number
operators.

\section{\label{sec:Luttinger-liquid-approach}Tomonaga-Luttinger liquid approach}

The universal behavior of this system can be found within a Tomonaga-Luttinger
liquid description~\cite{giamarchi_book}. In this paper, we are
interested in the phase diagram of the system at various densities
and interactions. First, we switch to a continuum description, $b_{a,i}\rightarrow b_{a}(x)$,
and express the operators $b_{a}(x)$ through a bosonization identity,
according to Haldane~\cite{Haldane,Cazalilla}: \begin{equation}
b_{a}(x)=[n+\Pi_{a}(x)]^{1/2}\sum_{m}e^{2mi\Theta_{a}(x)}e^{i\phi_{a}(x)},\end{equation}
 where the real-space density of each species is $n=\nu/a_{L}$ and
$a_{L}$ is the lattice constant. The lattice sites are at positions
$x=ia_{L}$. This expression is a phase-density representation of
the Bose operators, in which the square root of the density operator
has been written in an intricate way. The fields $\Pi_{1,2}(x)$ describe
the small amplitude and the long wave length density fluctuations.
The fields $\Theta_{1,2}(x)$ are given by $\Theta_{1,2}(x)=\pi nx+\theta_{1,2}(x)$,
where $\theta_{1,2}(x)=\pi\int^{x}dy\Pi_{1,2}(y)$. The fields $\phi_{1,2}(x)$
describe the phase, and are conjugate to the density fluctuations
$\Pi_{1,2}(x)$.

The contact interactions between the densities in \ref{eq:hamiltonian}
written in Haldane's representation generate an infinite series of
terms that contain $\exp(2m_{1}i(\pi nx+\theta_{1})+2m_{2}i(\pi nx+\theta_{2}))$,
where $m_{1}$ and $m_{2}$ are some integers. A term of this form
can only drive a phase transition, if the oscillatory part $2\pi m_{1}nx+2\pi m_{2}nx$
vanishes for all lattice sites. This leads to the requirement $m_{1}\nu+m_{2}\nu=m_{3}$,
with $m_{3}$ another integer~\cite{Commensurate}. As a further
requirement, small integers $m_{1}$ and $m_{2}$ are necessary, because
the scaling dimension of the term scales quadratically in $m_{1}$
and $m_{2}$.

For the range $0<\nu\leq1$, we find that there are three different
cases: unit-filling ($\nu=1$), half-filling ($\nu=1/2$), and non-commensurate
filling ($\nu\neq1$ and $\nu\neq1/2$). It can be checked, using
renormalization group arguments as below, that higher forms of commensurability
do not generate new phases, but that either phase separation or collapse
is reached first. Our numerical findings are consistent with this.

\emph{Non-commensurate filling.} The action of the system, assuming
a short-range spatial cut-off $r_{0}$, at non-commensurate filling
is given by~\cite{giamarchi_book,Cazalilla,supersolid_LM}: \begin{eqnarray}
S & = & \int d^{2}r[\sum_{j=1,2}\frac{1}{2\pi K}\left(\left(\partial_{v\tau}\theta_{j}\right)^{2}+\left(\partial_{x}\theta_{j}\right)^{2}\right)\nonumber \\
 &  & +\frac{U_{12}a_{L}}{\pi^{2}v\hbar}\partial_{x}\theta_{1}\partial_{x}\theta_{2}+\frac{2g_{\sigma}}{(2\pi r_{0})^{2}}\cos(2\theta_{1}-2\theta_{2})]\label{eq:Snoncomm}\end{eqnarray}
 The first line of the action is characterized by a Luttinger parameter
$K$ and a velocity $v$, contained in ${\bf r}=(v\tau,x)$. This
part of the action, without the coupling between the two fields $\theta_{a}(x)$,
generates a linear dispersion $\omega=v|k|$, where . $v$ should
therefore be interpreted as the phonon velocity. The Luttinger parameter
$K$ is a measure of the intra-species interaction $U$. We will be
interested in the regime $U\gtrsim t$, in which we have approximately
\cite{CazalillaTonks} \begin{eqnarray}
K & \approx & 1+\frac{8t}{U}\frac{\sin\pi\nu}{\pi}.\label{eq:K}\end{eqnarray}
 The velocity $v$ can also be related to the parameters of the underlying
Hubbard model by \begin{equation}
v\approx v_{F}(1-8t\nu\cos\pi\nu/U)\label{eq:velocity}\end{equation}
 where $v_{F}$ is the `Fermi velocity' of an identical system of
fermions, $v_{F}=2(a_{L}t/\hbar)\sin\pi\nu$, and $k_{F}$ is the
'Fermi wave vector', $k_{F}=\pi n$. Here, $\hbar$ is the Planck
constant.

The two fields $\theta_{a}(x)$ are coupled by the inter-species interaction.
The interaction term $U_{12}n_{1}n_{2}$ in the underlying Hubbard
model generates both the term containing $\partial_{x}\theta_{1}\partial_{x}\theta_{2}$,
as well as the backscattering term \cite{giamarchi_book,Commensurate}
containing $\cos(2\theta_{1}-2\theta_{2})$. The action $S$ is only
well-defined with a short-range cut-off $r_{0}$. It is proportional
to $1/n$. At this scale, $g_{\sigma}$ is approximately given by
\begin{eqnarray}
g_{\sigma} & = & U_{12}a_{L}/(v\hbar).\label{eq:gsigma}\end{eqnarray}

We diagonalize the quadratic part of the action by switching to the
symmetric and antisymmetric combinations $\theta_{S/A}=\frac{1}{\sqrt{2}}(\theta_{1}\pm\theta_{2})$.
For the two sectors we find \begin{eqnarray}
K_{S/A} & = & (1/K^{2}\pm U_{12}a_{L}/(v\hbar\pi K))^{-1/2}\label{eq:KSA}\end{eqnarray}
 as effective Luttinger parameters. To lowest order in $U_{12}$ this
gives $K_{S/A}\approx K\mp U_{12}a_{L}K^{2}/(2\pi v\hbar)$. The effective
velocities are $v_{S/A}=v\sqrt{1\pm U_{12}a_{L}K/(\pi v\hbar)}$.
Collapse (phase separation) of the superfluid phase is when $v_{S/A}$
is imaginary. We note that $K_{S}$ diverges when collapse (CL) is
approached, and that $K_{A}$ diverges as the system approaches phase
separation (PS).

The anti-symmetric sector contains the nonlinear backscattering term
$\cos(2\sqrt{2}\theta_{A})$. To study its effect, we use an RG approach.
We renormalize the short-range cut-off $r_{0}$ to a slightly larger
value, and correct for it at one-loop order. The resulting flow equations
are given by \cite{giamarchi_book}: \begin{eqnarray}
\frac{dg_{\sigma}}{dl} & = & (2-2K_{A})g_{\sigma}\label{eq:RG}\\
\frac{dK_{A}}{dl} & = & -\frac{g_{\sigma}^{2}}{2\pi^{2}}K_{A}^{3}\label{eq:RG_Ka}\end{eqnarray}
 The flow parameter $l$ is given by \begin{eqnarray}
l & = & \log_{e}\left(\frac{r_{0}'}{r_{0}}\right),\label{flowparameter}\end{eqnarray}
 where $r_{0}'$ is the new cut-off that has been created in the RG
process.

The flow equations \ref{eq:RG} and \ref{eq:RG_Ka} have two qualitatively
different fixed points: Either $g_{\sigma}$ diverges, which in turn
renormalizes $K_{A}$ to zero, or $g_{\sigma}$ is renormalized to
zero for finite $K_{A}=K_{A}^{*}$. In the latter case, the action
$S$ is quadratic in $\theta_{S}$ and $\theta_{A}$. For the parameter
$K_{S}$, we use the bare value given in Eq. \ref{eq:KSA}.

\begin{figure}
\includegraphics[width=5.4cm]{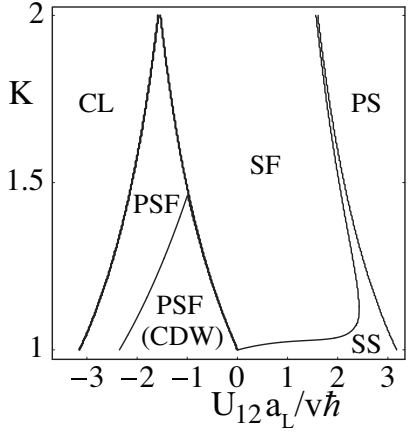}

\caption{\label{PDnoncomm} Phase diagram of a bosonic mixture at non-unit
and non-half-filling. For attractive interactions $U_{12}$ and $K<2$
the system can form a paired superfluid state, in the regime labeled
PSF and PSF(CDW). This phase can coexist with CDW order for weaker
interactions. For large repulsive (attractive) interactions $U_{12}$
the system phase separates (PS) (collapses (CL)). For the remaining
regime the system shows single particle superfluidity (SF). This can
coexist with CDW order, resulting in a quasi-supersolid (SS) regime.}

\end{figure}

As mentioned in the introduction, we can determine the phase diagram
by studying the long-range scaling behavior of correlation functions,
$\langle O^{\dagger}(x)O(y)\rangle$, of various order parameters
$O(x)$. In particular, the single-particle superfluid order parameter
is $O_{SF}=b_{a}(x)$ with $a=1,2$. The CDW order is related to the
$2k_{F}$ wavevector component of the density operator, $O_{CDW}=n_{a}$.
PSF is described by $O_{PSF}=b_{1}(x)b_{2}(x)$, and CFSF by $O_{CFSF}=b_{1}^{\dagger}(x)b_{2}(x)$.
In the homogeneous system, it suffices to study \begin{eqnarray}
G(x) & = & \langle b_{a}^{\dagger}(x)b_{a}(0)\rangle,a=1,2\label{eq:spG}\\
R_{n,a}(x) & = & \langle n_{a}(x)n_{a}(0)\rangle,a=1,2\label{eq:denscorr}\\
R_{S}(x) & = & \langle b_{1}^{\dagger}(x)b_{2}^{\dagger}(x)b_{1}(0)b_{2}(0)\rangle\label{eq:pcorr}\\
R_{A}(x) & = & \langle b_{1}^{\dagger}(x)b_{2}(x)b_{1}(0)b_{2}^{\dagger}(0)\rangle.\label{eq:acorr}\end{eqnarray}
 We find that away from collapse (CL) and phase separation (PS), the
correlation functions scale either algebraically or exponentially.
For algebraic scaling, we have \begin{eqnarray}
G(x) & \sim & |x|^{\alpha_{SF}-2},\nonumber \\
 &  & \alpha_{SF}=2-1/(4K_{S})-1/(4K_{A})\label{eq:alphaSF}\\
R_{n,a}(x) & \sim & \cos(2k_{F}x)|x|^{\alpha_{CDW}-2},\nonumber \\
 &  & \alpha_{CDW}=2-K_{S}-K_{A}\label{eq:alphaCDW}\\
R_{S}(x) & \sim & |x|^{\alpha_{PSF}-2},\alpha_{PSF}=2-1/K_{S}\label{eq:alphaPSF}\\
R_{A}(x) & \sim & |x|^{\alpha_{CFSF}-2},\alpha_{CFSF}=2-1/K_{A}.\label{eq:alphaCFSF}\end{eqnarray}
 where the scaling exponents $\alpha_{O}$ are determined by $K_{S}$
and $K_{A}$ after the RG flow. For the case that $g_{\sigma}$ diverges
in Eqs. \ref{eq:RG} and \ref{eq:RG_Ka} and $K_{A}$ is undefined,
these expressions can still be used. We set $K_{A}$ to zero, and
find that $\alpha_{CDW}$ and $\alpha_{PSF}$ are well defined. Hence
$R_{n,a}$ and $R_{S}$ still show algebraic scaling. On the other
hand, $\alpha_{CFSF}$ and $\alpha_{SF}$ become $-\infty$ and $G$
and $R_{A}$ scale exponentially.

We can identify regimes where different scaling exponents are positive
based on the relationship between the scaling exponents and $K_{S/A}$
after the flow. This determines the different quasi-long range orders
that are present. The resulting phase diagram is shown in Fig. \ref{PDnoncomm},
as a function $K$ and $U_{12}a_{L}/(v\hbar)$, as appearing in the
action in Eq.~\ref{eq:Snoncomm}. These two parameters determine
the initial values of the flow equations through equations \ref{eq:KSA}
and \ref{eq:gsigma}.

We can estimate the phase boundary between PSF and SF. For small $U_{12}a_{L}/(v\hbar)$
this boundary is near the point $K_{A}=1$ and $g_{\sigma}=0$. For
that limit, Eq.~(\ref{eq:RG_Ka}) can be linearized to \begin{eqnarray}
\frac{dK_{A}}{dl} & = & -\frac{g_{\sigma}^{2}}{2\pi^{2}}\label{eq:RG_Ka2}\end{eqnarray}
 and the expression $A=\pi^{2}(1-K_{A})^{2}-g_{\sigma}^{2}/4$ becomes
an invariant of the flow. From the properties of the RG flow of a
Berezinskii-Kosterlitz-Thouless transition (see e.g. \cite{giamarchi_book,BKT}),
the phase boundary is given by $A=0$ and $g_{\sigma}<0$. Using the
expressions of $K_{A}$ and $v$ in terms of the Hubbard parameters,
we estimate the critical interaction $U_{12}$ for PSF to occur at
\begin{eqnarray}
\frac{U_{12}}{U}\Bigg|_{c} & = & -32\frac{t^{2}}{U^{2}}\sin^{2}(\pi\nu).\label{eq:criticalPSF}\end{eqnarray}
 The phase boundary between supersolid (SS) and SF has been derived
in Ref.~\cite{supersolid_LM}.

\begin{figure}
\includegraphics[width=5.4cm]{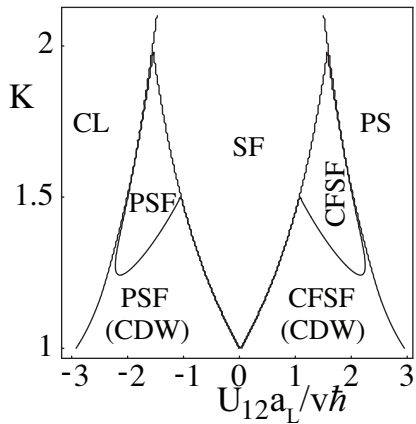}

\caption{\label{PD_half} Phase diagram of a bosonic mixture at half-filling.
In addition to the phases that appear in Fig. \ref{PDnoncomm}, the
system now develops a counterflow superfluid (CFSF) phase, which can
coexist with CDW order. }

\end{figure}

\textit{Half-filling.} In the case of half-filling, another non-linear
term has to be introduced in the action \begin{eqnarray}
S_{uk} & = & \frac{2g_{uk}}{(2\pi r_{0})^{2}}\int d^{2}r\cos(2\theta_{1}+2\theta_{2}).\label{Suk}\end{eqnarray}
 This term describes Umklapp scattering. At the initial cut-off $r_{0}\sim1/n$,
$g_{uk}$ is approximately given by $U_{12}a_{L}/v$. In addition
to the RG flow in the antisymmetric sector we now also have \begin{eqnarray}
\frac{dg_{uk}}{dl} & = & (2-2K_{S})g_{uk}\label{eq:RG_half}\\
\frac{dK_{S}}{dl} & = & -\frac{g_{uk}^{2}}{2\pi^{2}}K_{S}^{3}\end{eqnarray}
 in the symmetric sector. Proceeding along the same lines as for the
non-commensurate case, we find the phase diagram shown in Fig. \ref{PD_half}.

We estimate the SF-CFSF phase boundary in the same way as the PSF-SF
boundary. We find \begin{eqnarray}
\frac{U_{12}}{U}\Bigg|_{c} & = & 32\frac{t^{2}}{U^{2}}\sin^{2}(\pi\nu).\label{eq:criticalCFSF}\end{eqnarray}

\textit{Unit-filling.} At unit-filling we have to introduce a term
of the form \begin{eqnarray}
S_{1} & = & \frac{2g_{1}}{(2\pi r_{0})^{2}}\int d^{2}r\left(\cos(2\theta_{1})+\cos(2\theta_{2})\right).\label{S1}\end{eqnarray}
 The resulting RG flow for this system is given by \begin{eqnarray}
\frac{dg_{uk}}{dl} & = & (2-2K_{S})g_{uk}+\alpha_{3}\frac{g_{1}^{2}(K_{A}-K_{S})}{2\pi}\label{RG_unit}\\
\frac{dg_{\sigma}}{dl} & = & (2-2K_{A})g_{\sigma}+\alpha_{3}\frac{g_{1}^{2}(K_{S}-K_{A})}{2\pi}\\
\frac{dg_{1}}{dl} & = & (2-\frac{K_{S}+K_{A}}{2}+\alpha_{3}\frac{g_{uk}K_{S}+g_{\sigma}K_{A}}{\pi})g_{1}\\
\frac{dK_{A}}{dl} & = & -\frac{g_{\sigma}^{2}}{2\pi^{2}}K_{A}^{3}-\frac{g_{1}^{2}}{16\pi^{2}}(K_{S}+K_{A})K_{A}^{2}\\
\frac{dK_{S}}{dl} & = & -\frac{g_{uk}^{2}}{2\pi^{2}}K_{S}^{3}-\frac{g_{1}^{2}}{16\pi^{2}}(K_{S}+K_{A})K_{S}^{2}\end{eqnarray}
 where $\alpha_{3}$ is some non-universal parameter \cite{SG}.
The behavior of this set of equations depends strongly on the initial
value of $g_{1}$. For small values of $g_{1}$, four phases can be
stable: Single-particle superfluidity, CFSF, PSF and a Mott phase.
For large values only single-particle SF and MI are stable. We determine
with our numerical approach, that the Hubbard model falls into the
second category, i.e. there is only a single-particle SF and a Mott
state at unit-filling.

Having established the universal behavior of the system from Tomonaga-Luttinger
liquid theory, we now want to connect the phase diagram with the parameters
in the Hubbard model. The expressions \ref{eq:K} and \ref{eq:velocity},
which relate the Luttinger parameter $K$ and the velocity $v$ to
microscopic parameters of the Hubbard model, are only approximate,
no full analytic expression is known. In addition, only some phase
boundaries are predicted reliably, because we use perturbative RG
in the $g_{\sigma}$. We expect that the analytic calculation only
predicts the general structure of the phase diagram, as well as the
decay behavior of the correlation functions. To obtain the phase diagram
in terms of the parameters in the Hubbard model, we need to use numerical
methods. The next section describes the numerical determination of
the phase diagram.

\section{\label{sec:Numerical-Approach}Numerical Approach}

We use the time-evolving-block-decimation (TEBD) method \cite{Vidal}
to study our discrete one-dimensional two-species Hubbard Hamiltonian.
With this method, explained in Appendix \ref{sec:The-effective-one-species},
we obtain an approximate ground state solution. We consider $N$ lattice
sites with hard-wall boundary conditions and express the Hubbard parameters
in units of the intra-species interaction $U$. The number of sites
$N$ is equal to 80, unless otherwise noted. In our numerical analysis,
we limit the particle number on each site and each species to two
for filling $\nu\leqslant0.8$ and four otherwise. Once we obtain
the ground state, we calculate the energy, density distributions,
correlation functions, and the structure factor to identify the quasi-long
range order and other properties of the ground state. %
\begin{table}
\begin{tabular}{|c|c|c|c|}
\hline 
 & $R_{S}(x)$  & $R_{A}(x)$  & $G(x)$\tabularnewline
\hline
\hline 
MI  & Exp.  & Exp.  & Exp.\tabularnewline
\hline 
SF  & Alg.  & Alg.  & Alg. \tabularnewline
\hline 
CFSF  & Exp.  & Alg.  & Exp. \tabularnewline
\hline 
PSF  & Alg.  & Exp.  & Exp.\tabularnewline
\hline 
CL/PS  & \multicolumn{3}{c|}{$R_{S}(x)$, $R_{A}(x)$ undefined}\tabularnewline
\hline
\end{tabular}

\caption{\label{tab:table}Definitions of Mott insulator (MI), superfluid (SF),
counterflow superfluid (CFSF) and paired superfluid (PSF) orders in
terms of the long-range behavior of the correlation functions $R_{S}(x)$,
$R_{A}(x)$, and $G(x)$ . Each of these can either show algebraic
(Alg.) or exponential (Exp.) decay when the system is away from collapse
(CL) or phase separation (PS). From the Tomonaga-Luttinger liquid
theory, $R_{S}(x)$ and $R_{A}(x)$ approach a constant (or $K_{S/A}$
diverges) when the system approaches CL/PS regime. For the numerical
calculation in the CL/PS regimes, the behavior of the correlation
functions is inconclusive and we assign the phase from additional
observables as discussed in the text. }

\end{table}

\begin{figure}
\includegraphics[width=8cm]{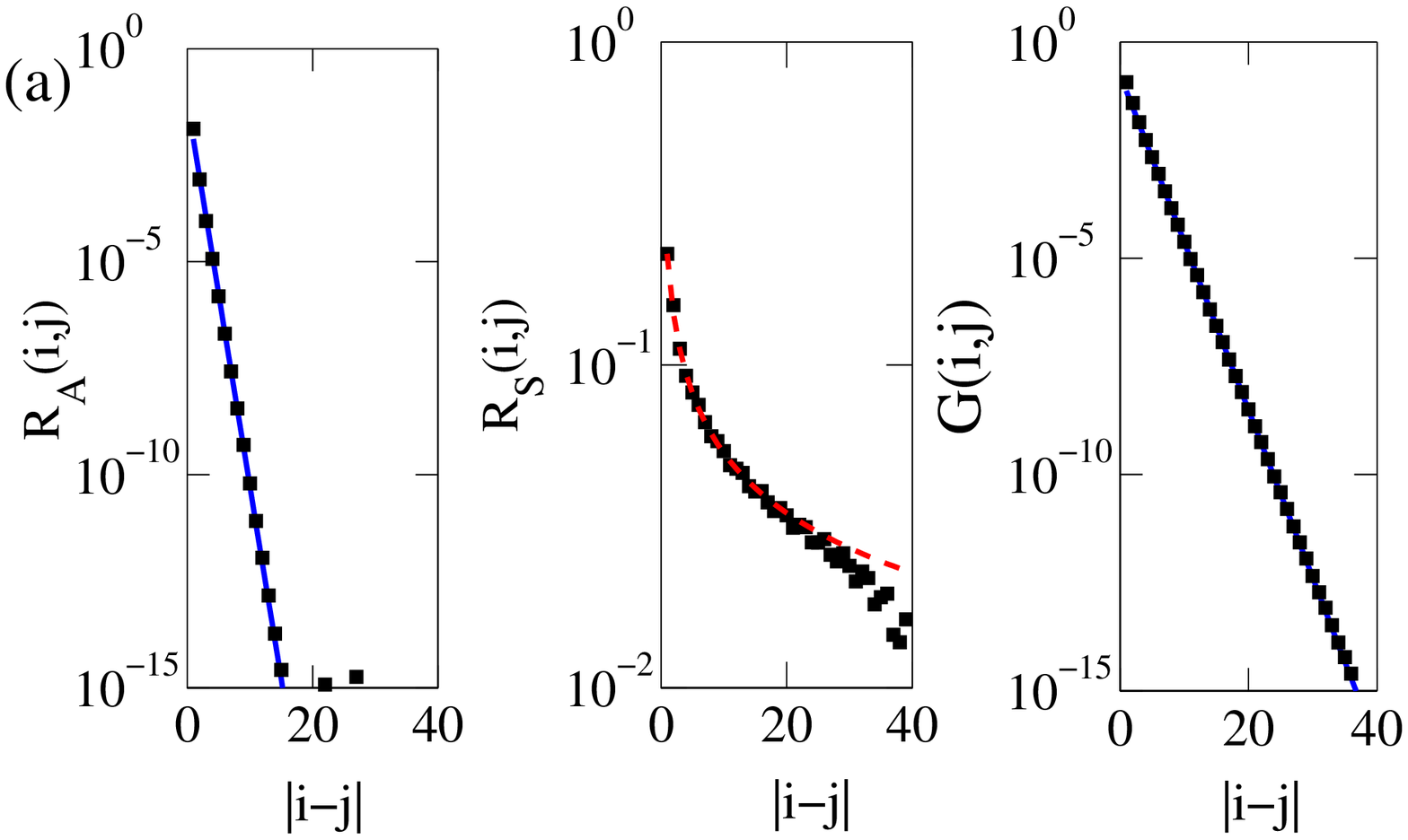}

\includegraphics[width=8cm]{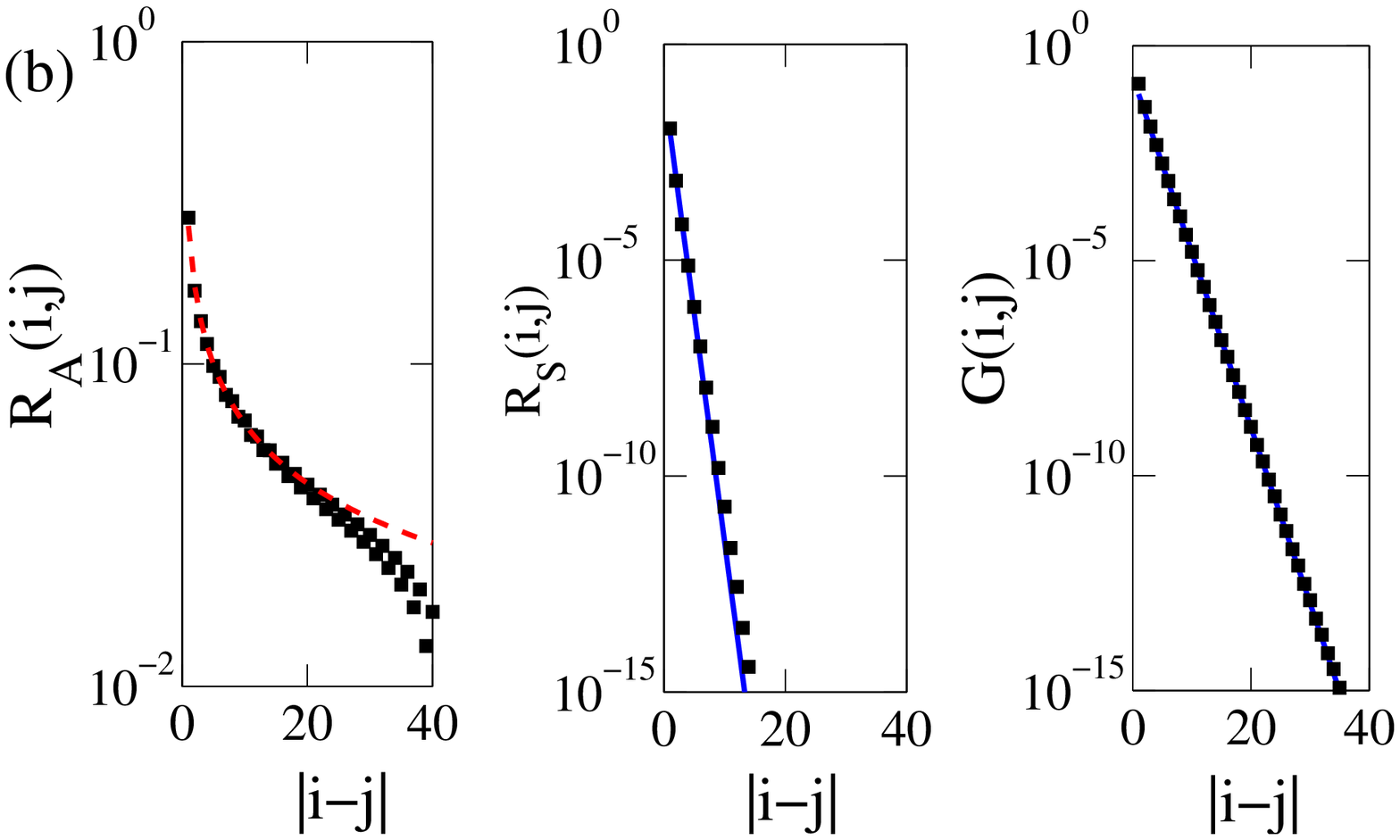}

\caption{\label{fig:correlations} The correlation functions $R_{A}$, $R_{S}$,
and $G$ on a logarithmic scale as a function of distance $|i-j|$.
The index $i$ is $40$, the center of the 80 lattice sites. The squares
are the numerical data. The blue lines are exponential fits to the
data and red dotted lines are algebraic fits. Note that the scale
of the vertical axis of the graphs differs by orders of magnitude.
In (a), we show an example for the paired superfluid phase at $\nu$=0.3,
$t=0.02U$, and $U_{12}=-0.16U$. $R_{A}$ decays exponentially and
$R_{S}$ decays algebraically. The single-particle correlation function
decays exponentially, implying the absence of single-particle superfluidity.
In (b), we show an example for the counterflow superfluid phase at
$\nu=0.5$, $t=0.02U$, and $U_{12}=0.2U$. The anti-pair correlation
function $R_{A}$ decays algebraically, while the pair correlation
function decays exponentially. Single-particle superfluidity is again
absent. The algebraic fits deviate from the data around $|i-j|\approx40$,
due to the boundary conditions of our numerical calculations.}

\end{figure}

For example, to determine whether a SF, PSF, or CFSF is present, we
study the decay behavior of the correlation functions, $G(x)$, $R_{A}(x)$,
and $R_{S}(x)$, defined in Eqs.~\ref{eq:spG},~\ref{eq:acorr},
and ~\ref{eq:pcorr}, respectively. If both $R_{A}$ and $R_{S}$
decay algebraically, the system is in a single-particle superfluid
(SF) state. If both are exponential, the system is in a Mott insulator(MI)
state. If $R_{S}$ or $R_{A}$ decays algebraically, the system is
in the PSF or CFSF state, respectively. These relationships are summarized
in Table \ref{tab:table}.

In Fig.~\ref{fig:correlations}(a) and (b), we show the decay behavior
of the correlation functions in the PSF and CFSF phase, respectively.
As the Hamiltonian is discrete, the correlation functions are calculated
as discrete functions: $G(i,j)=\langle b_{a,i}^{\dagger}b_{a,j}\rangle$,
$R_{S}(i,j)=\langle b_{1,i}^{\dagger}b_{2,i}^{\dagger}b_{1,j}b_{2,j}\rangle$,
and $R_{A}(i,j)=\langle b_{1,i}^{\dagger}b_{2,i}b_{1,j}b_{2,j}^{\dagger}\rangle$.
For the PSF phase, $R_{A}(i,j)$ decays exponentially, while $R_{S}$
decays algebraically. It is also worthwhile to notice that the single-particle
Green's function decays \emph{exponentially}, implying the absence
of single-particle superfluidity. For the CFSF phase, $R_{A}$ decays
algebraically while $R_{S}$ decays exponentially. Single-particle
superfluidity is again absent.

\begin{figure}
\includegraphics[width=8.5cm]{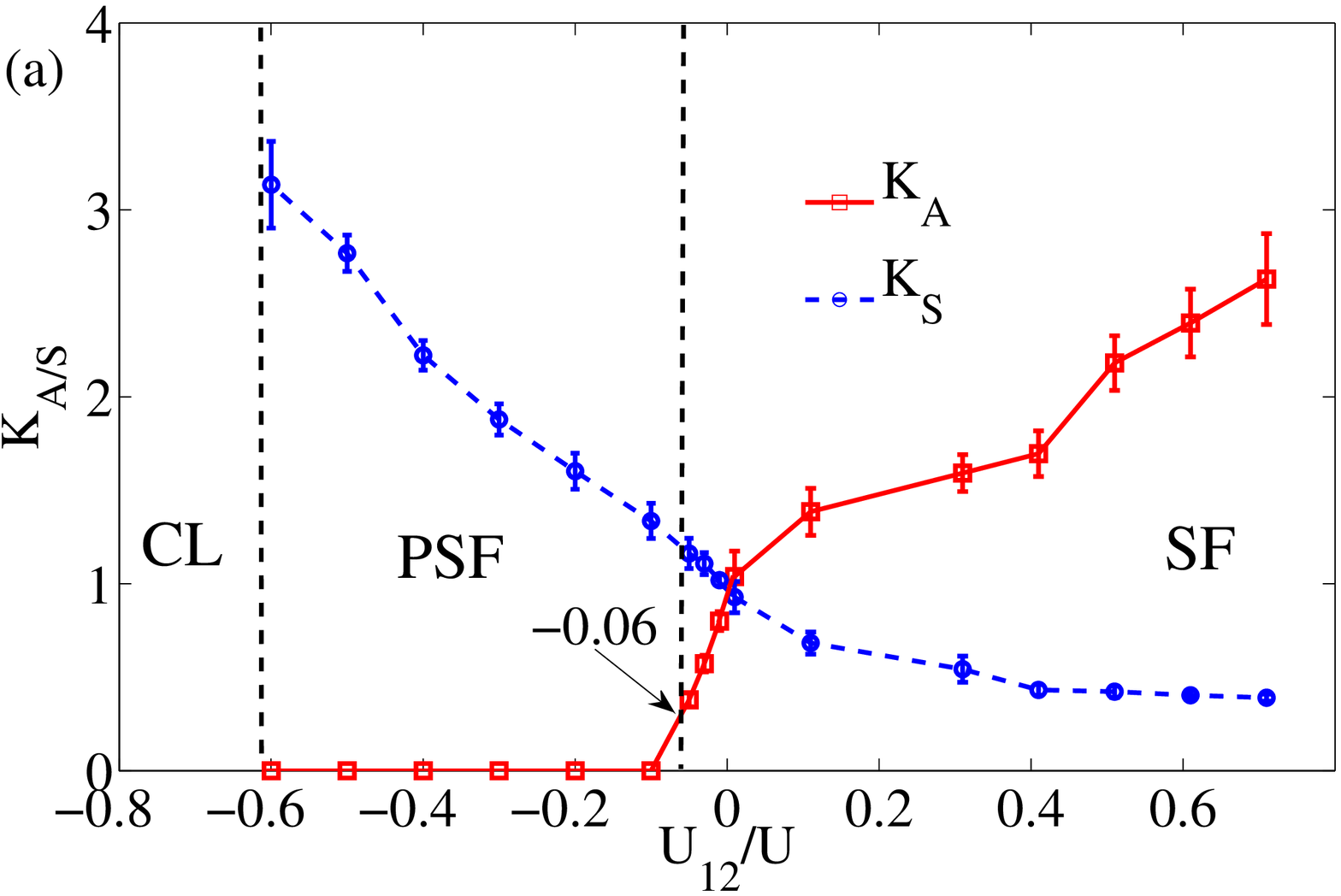}

\includegraphics[width=8.5cm]{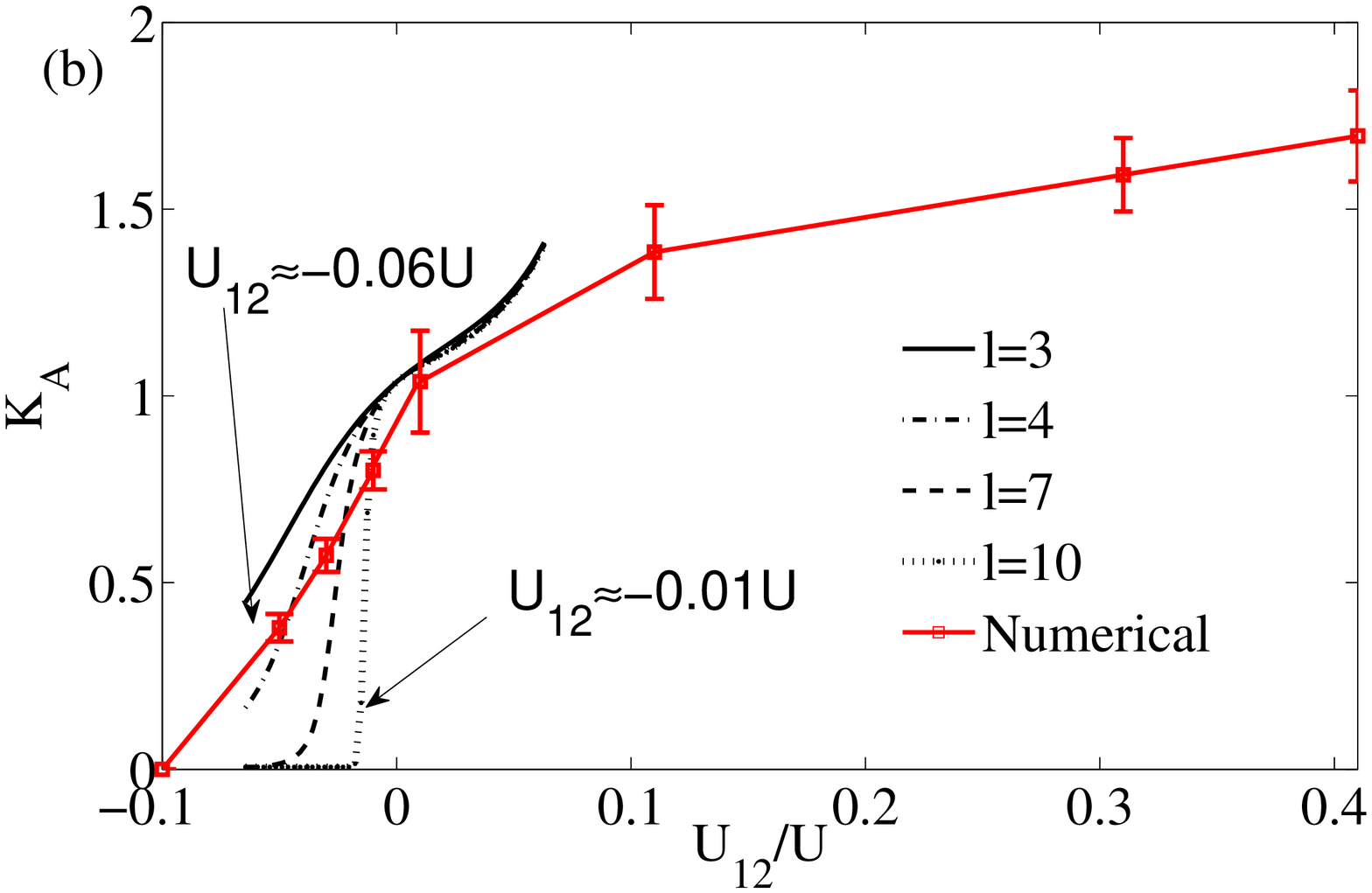}

\caption{\label{fig:K112} (a) $K_{S}$ and $K_{A}$ as a function of $U_{12}$
as extracted from the fit of the correlation functions, $R_{S}$ and
$R_{A}$. The filling $\nu$ is 0.7 and $t/U$ is 0.02. Around $U_{12}/U\approx-0.06$,
the anti-pair correlation function changes from algebraic to exponential
decay. This corresponds to the transition from the PSF to SF phase.
When $R_{A}$ decays exponentially, $K_{A}$ is formally set to zero.
For $K_{a}+K_{s}\lesssim2$, the system has CDW order. Error bars
are one standard deviation uncertainties obtained from the power-law
fit to the numerical data. (b) A comparison of $K_{A}$ obtained from
our RG and TEBD calculations. The red square connected by lines are
the TEBD results while all other lines are determined from the RG
flow with flow parameter $l=3,4,7$, and 10, where $l$ is defined
in Eq.~\ref{flowparameter}. The error bars are as in panel (a).
The PSF-to-SF transition obtained from TEBD is around $U_{12}/U\thickapprox-0.06$,
while the RG calculation shows that for $l=10$, the transition occurs
near $U_{12}/U\thickapprox-0.01$. We interpret the regime between
$U_{12}/U\thickapprox-0.06$ and $-0.01$ the cross-over region.}

\end{figure}

\emph{Behavior of $K_{S}$ and $K_{A}$}: We study the decay behavior
of $R_{S}$ and $R_{A}$ in more detail. Using the fit function, $c\cdot|i-j|^{\alpha-2}$,
where $c$ and $\alpha$ are the fitting parameters, we obtain the
power-law exponent $\alpha$ and, hence, the Luttinger parameters
$K_{S}$ and $K_{A}$ based on Eqs.~\ref{eq:alphaPSF} and \ref{eq:alphaCFSF}.
In Fig.~\ref{fig:K112}(a), we show these $K_{S}$ and $K_{A}$ as
a function of $U_{12}$, for non-commensurate filling. A Luttinger
parameter is formally set to zero when its correlation function decays
exponentially.

For $U_{12}<-0.06U$, $R_{A}$ decays exponentially, while for $U_{12}>-0.06U$,
$R_{A}$ decays algebraically, and $K_{A}$. increases as $U_{12}$
increases. The system undergoes a PSF to SF transition at $U_{12}=-0.06U$.
On the other hand, $K_{S}$ decreases monotonically for $U_{12}>-0.6U$.
For $U_{12}<-0.6U$ the numerics failed to converge to a homogeneous
state. This indicates that the system collapses, and we therefore
cannot extract a Luttinger liquid parameter. We can observe charge
density wave (CDW) order for a range of $U_{12}/U$ in Fig.~\ref{fig:K112}.
According to Eq.~\ref{eq:alphaCDW}, this order exists when $K_{S}+K_{A}<2$.
In fact, it co-exists with the SF, PSF or CFSF order. At half-filling,
$K_{S}$ will go to zero at a critical, positive value of $U_{12}$.
This indicates the transition from the SF to CFSF phase.

\emph{Finite-size effect\label{Finite-size-effect:-The}}: The behavior
of $K_{A/S}$ stated above is affected by the size of the system.
Finite size effects can 'smooth out' a sudden change in $K_{A/S}$
at the phase transition. This effect can be estimated from the RG
flow calculation by integrating Eqs.~\ref{eq:RG} and \ref{eq:RG_Ka}
out to a finite value $l$ rather than to infinity. In Fig.~\ref{fig:K112}(b),
we show an example of a finite-$l$ RG calculation in the vicinity
of the PSF-to-SF transition. We see that as $l$ increases, $K_{A}$
dramatically changes for the attractive $U_{12}$. In the limit of
$l\rightarrow\infty$, $K_{A}$ becomes discontinuous and 'jumps'
from 0 to 1 at $U_{12}\thickapprox-0.01U$. This is where the PSF-to-SF
transition occurs. This transition is a Berezinskii-Kosterlitz-Thouless
transition \cite{BKT,giamarchi_book}. In order to compare the RG
result with our TEBD result, we associate the system size $N$ with
the flow parameter $l$, based on the relation in Eq.~\ref{flowparameter}.
The cut-off $r_{0}$ is the lattice constant $a_{L}$ and $r_{0}'=Na_{L}$.
For $N=80$ we have $l=4.4$ and we find that the RG and TEBD are
in good agreement. The regime between $U_{12}/U\approx-0.06$ and
$-0.01$ is a cross-over regime due to the finite size of the system.

\begin{figure*}
\includegraphics[clip,width=15cm]{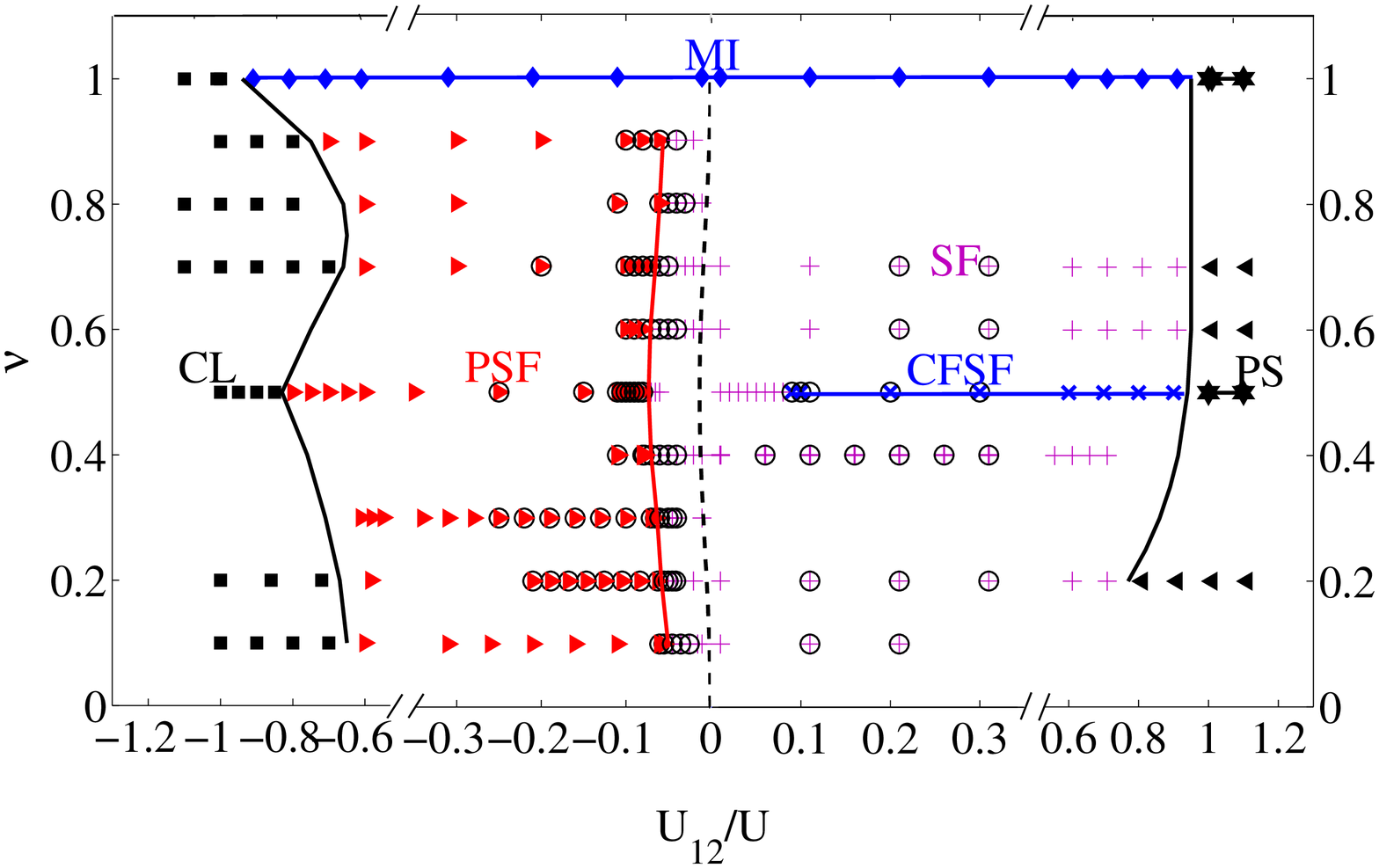}\includegraphics[bb=14bp 14bp 241bp 0bp,width=2.5cm]{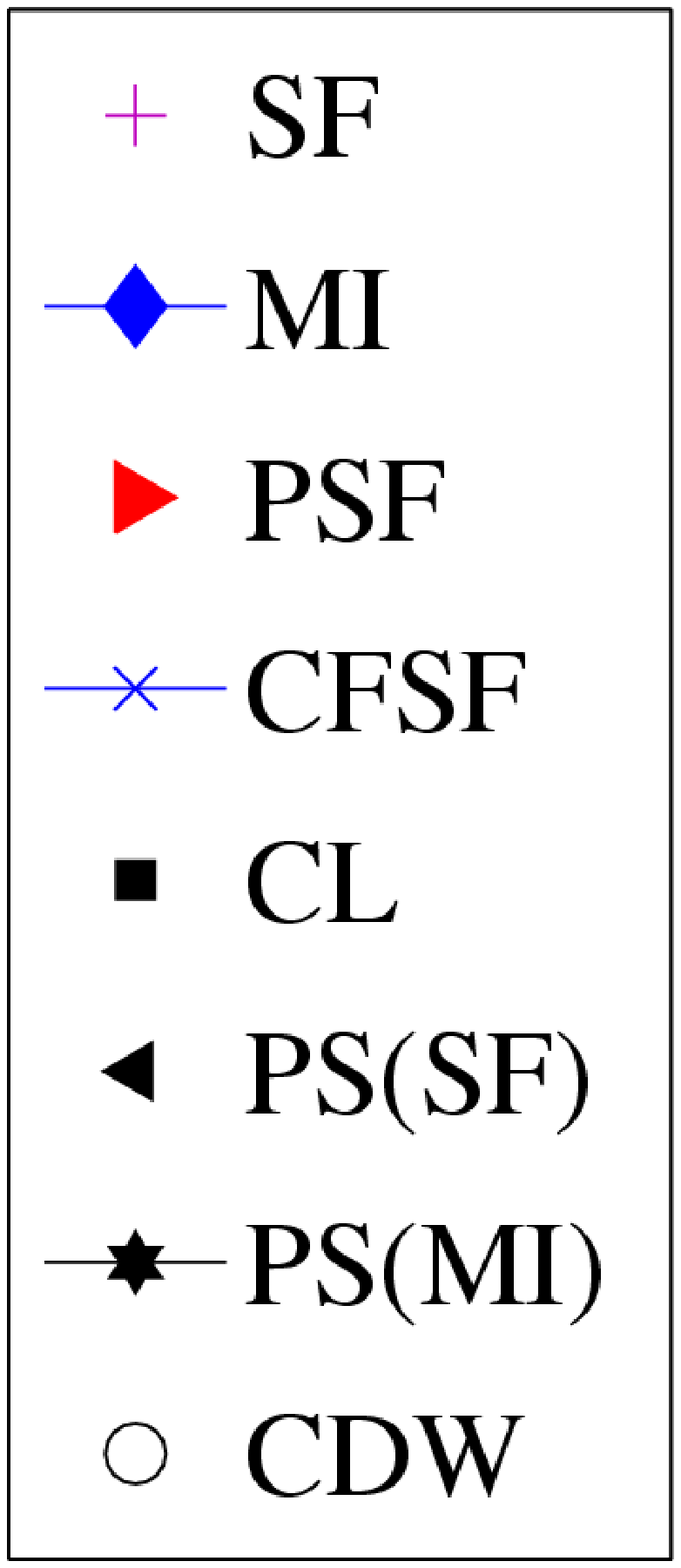}

\caption{\label{fig:phasedia}Phase diagram for a homogeneous system with 80
sites and the hopping parameter $t=0.02$U as a function of filling
$\nu$ and inter-species interaction $U_{12}/U$. The horizontal axis
shows three disconnected regions in $U_{12}/U$. The solid lines are
the estimated phase boundaries based on the TEBD results and the dotted
line is the PSF-to-SF phase boundary predicted by our RG calculation
(see Eq.~\ref{eq:criticalPSF}). For attractive interaction $U_{12}\lesssim-0.06U$,
the system forms a paired-superfluid (PSF). The state collapses(CL)
for $U_{12}\lesssim-0.7U$. For $U_{12}\gtrsim-0.06$ and $U_{12}\lesssim U$
the system shows single-particle superfluidity (SF). The system phase-separates
(PS) for $U_{12}\gtrsim1$ and forms two single-particle superfluids
(SF). Open circles are the points where $K_{S}+K_{A}<2$ and charge
density wave (CDW) order coexists with a superfluid phase (SF,PSF,
or CFSF). At half and unit filling there exist special phases. For
repulsive interaction $U_{12}\gtrsim0.08U$ and half-filling, the
system forms a counterflow superfluid (CFSF). For unit filling, we
find a Mott-Insulator (MI) phase for interactions $|U_{12}|\lesssim U$.
Finally, in the PS region at half- and unit-filling, the system forms
two individual MI states. }

\end{figure*}

\emph{Collapse and phase separation:} For large $|U_{12}|$, the system
approaches collapse or phase separation. According to Tomonaga-Luttinger
liquid theory, $K_{S}\rightarrow\infty$ as the system approaches
collapse and $K_{A}\rightarrow\infty$ as the system approaches phase
separation. As seen in Fig.~\ref{fig:K112}, we indeed find such
a tendency in our TEBD calculations. For $U_{12}>0.8U$ (not shown),
$K_{A}$ increases rapidly to values around 10, indicating a possible
phase separation. For $U_{12}<-0.6U$, due to the slow decay of the
correlation function $R_{S}$ and the finite-size of our system, we
are unable to extract an accurate $K_{S}$ from the numerical result.
On the other hand, we observe a peaked density distribution for $U_{12}<-0.6U$,
indicating a collapse. In the phase separation regime, $G(x)$ has
algebraic decay except for $\nu=0.5$ or $1$, where it has exponential
decay. An algebraic decay implies two spatially-separated single-species
superfluids while the exponential decay implies two spatially-separated
Mott insulators.\cite{Mishra}.

\subsection{\label{sec:The-phase-diagrams}Phase diagram}

\begin{figure*}[t]
 \includegraphics[width=13.5cm]{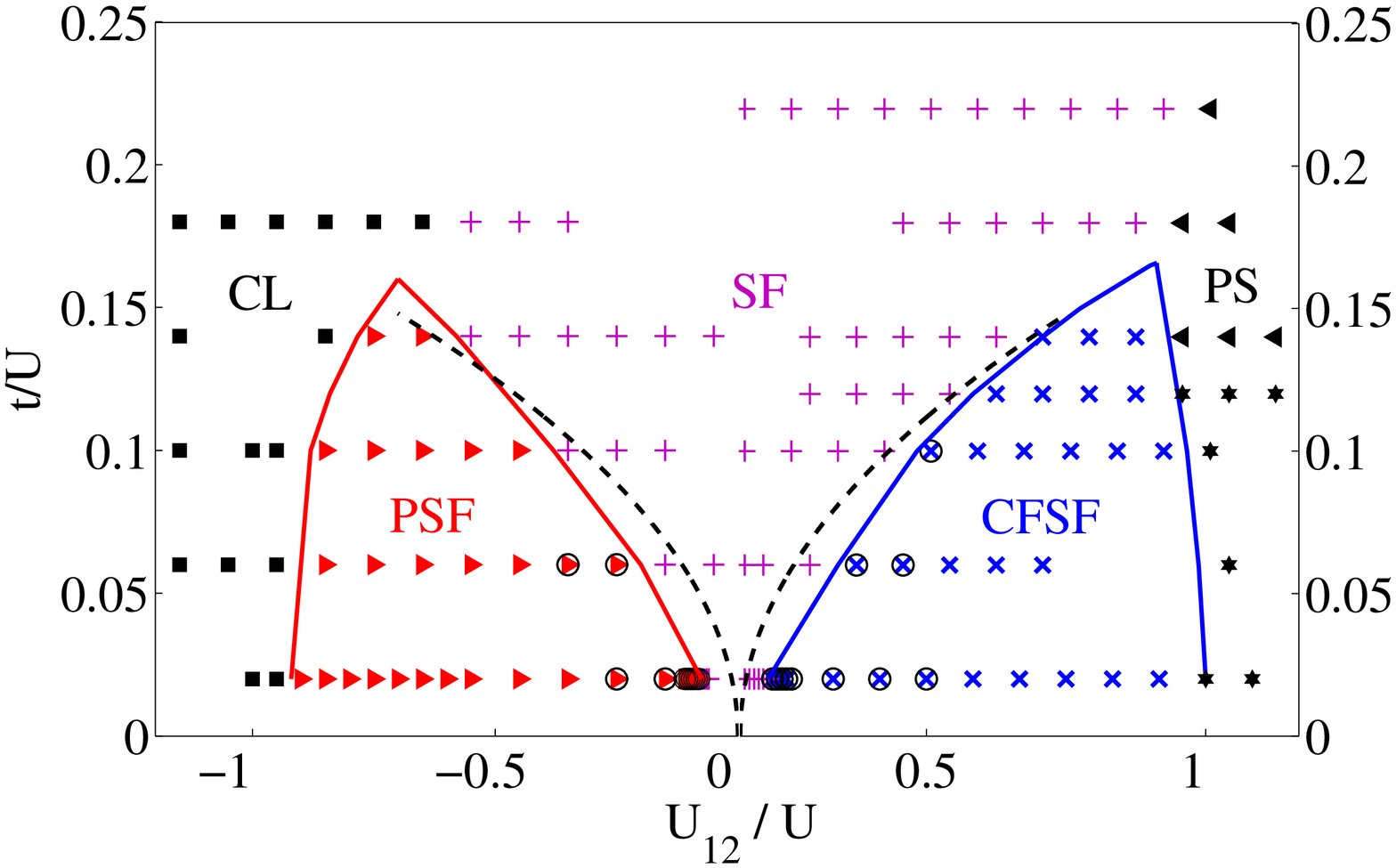}\includegraphics[width=2.5cm]{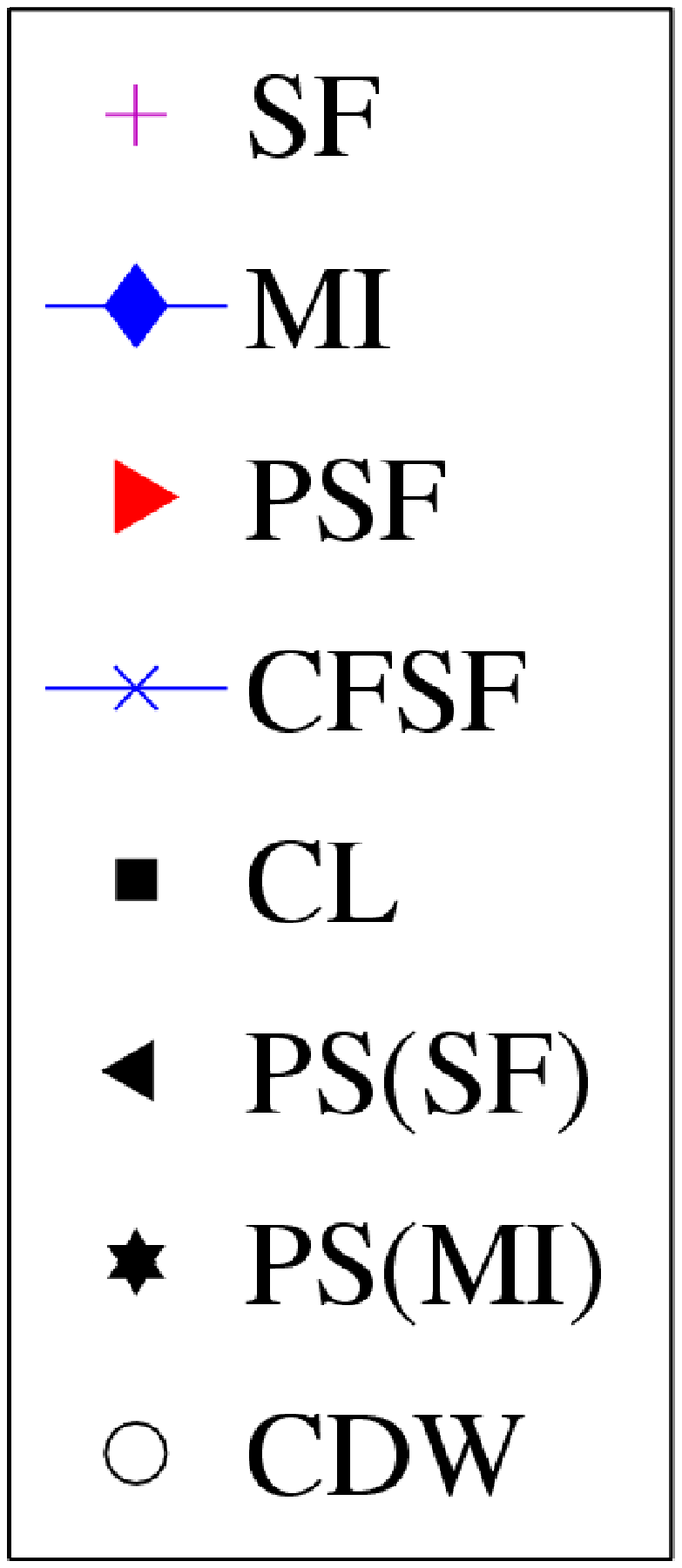}

\caption{\label{fig:The-phase-diagram at half filling}Phase diagram at half-filling
as a function of $U_{12}/U$ and $t/U$. The solid lines are estimated
phase boundaries from the TEBD calculation and the dotted lines are
the phase boundaries predicted by the RG calculation (see Eqs.~\ref{eq:criticalPSF}
and~\ref{eq:criticalCFSF}). For large repulsive interaction, the
system phase separates (PS) and for large attractive interaction,
the system collapses (CL). For moderate interactions and for $t/U\lesssim0.2$,
the system shows paired superfluidity (PSF) on the attractive side
and counterflow superfluidity on the repulsive side. Both PSF and
CFSF can coexist with charge density wave (CDW) order when $t\lesssim0.1U$. }

\end{figure*}

We study the phase diagram as a function of filling $\nu$ and parameters
of the Hubbard Hamiltonian. Assuming a positive $U$, the system can
be fully characterized in terms of $\nu$, $t/U$, and $U_{12}/U$.
Our results are shown in Fig.~\ref{fig:phasedia} for a fixed hopping
parameter and in Fig.~\ref{fig:The-phase-diagram at half filling}
for half filling.

\subsubsection{Phase diagram at a fixed hopping parameter}

In Fig.~\ref{fig:phasedia} we show the phase diagram for filling
fractions between 0 and 1 and the interaction $U_{12}/U$ between
-1.1 and 1.1. The symbols correspond to numerical data points at which
the phases have been characterized. Different markers represent the
different orders. The orders are determined from the decay behavior
of the three correlation functions $R_{A}$, $R_{S}$, and $G$.

For weak attractive inter-species interaction, $-0.06<U_{12}/U<0$,
the system is in a SF state. As $U_{12}$ grows more attractive, paired
superfluidity (PSF) occurs. The critical $U_{12}$ is largest, $\sim-0.08U$,
at half-filling and gradually decreases away from half-filling. This
phase boundary differs from that predicted by our RG calculation (Eq.
\ref{eq:criticalPSF}), plotted as the dotted line in Fig.~\ref{fig:phasedia}.
This discrepancy is the result of the finite-size effect discussed
in Fig.~\ref{fig:K112}(b). In the SF to PSF cross-over regime, charge
density wave (CDW) order can coexist. According to the phase diagram
Fig.~\ref{PDnoncomm}, for attractive interaction, CDW order can
co-exist only with PSF order. In our numerical work, we observed the
CDW order slightly outside the numerical phase boundary of PSF but
within the RG phase boundary of PSF. The sub-regime where CDW and
PSF coexist ends when $U_{12}/U\lesssim-0.4$. When the inter-species
attraction is comparable to the intra-species repulsion, $U_{12}\lesssim-U$,
the system collapses (CL) and no long-range order is present.

For repulsive inter-species interaction and $U_{12}<U$, the system
is in a SF state for all non-commensurate fillings. Within the SF
regime, there is a smaller parameter region where CDW order coexist
with the SF order. This subregime is a quasi-supersolid regime. The
boundary between a normal superfluid and a quasi-supersolid is estimated
by RG calculation in Ref.~\cite{supersolid_LM}. At half-filling,
counterflow superfluidity (CFSF) occurs when $0.08\lesssim U_{12}/U\lesssim1$.
Within the CFSF regime, the CDW order can coexist, forming a quasi-supersolid
of anti-pairs. It also worthwhile to point out that at half-filling,
CDW order only exists within the PSF and CFSF regimes.

At unit filling, our numerical results do not show evidence of PSF
or CFSF for any $U_{12}$. We find a Mott insulator (MI) state for
$|U_{12}|<U$.

\subsubsection{Phase diagram at half-filling}

In Fig.~\ref{fig:The-phase-diagram at half filling}, we show the
phase diagram at half filling as a function of $U_{12}/U$ and $t/U$.
From this diagram, we find that the border between PSF and SF and
the border between PSF and CL approach each other as $t$ increases.
Similarly, the border between the CFSF and SF and the border between
CFSF and PS approach each other. In fact, the PSF and CFSF phases
end around $t\thicksim0.16U$. Within the PSF and CFSF regimes, CDW
order can co-exist. In the phase separated regime, the separated single-species
ensembles form two individual Mott insulating states for $t\lesssim0.14U$
and two individual SF states for $t>0.14U$.

We can compare this phase diagram with the half-filling phase diagram
in Fig.~\ref{PD_half} obtained from Tomonaga-Luttinger liquid theory.
Especially, we can compare the location of the phase boundary between
SF and PSF(CFSF). To do so, we plot the RG phase boundaries, described
by Eqs.~\ref{eq:criticalPSF} and \ref{eq:criticalCFSF}, onto our
phase diagram. The area near the two boundaries is interpreted as
the cross-over regime where finite-size effects modify the phase boundary.

\subsection{\label{sec:Experimental-realization-and} Realization and detection}

Having established the phase diagram for the homogeneous system, we
now discuss how to realize and detect the PSF and CFSF phases. First,
we need to modify the Hubbard Hamiltonian in Eq.~\ref{eq:hamiltonian}
because in any ultra-cold atom experiment an additional trapping potential
is present. We add a harmonic potential, $\Omega(j-j_{c})^{2}(n_{1,j}+n_{2,j})$,
where $j$ is the site index and $j_{c}$ is the index at the center
of the system. The TEBD method is used to find the ground state. We
consider a system of 80 lattice sites and adjust the total number
of particles and the trap frequency so that the number of particles
is negligible at the edge of the lattice.

We again determine the orders of the system by studying the correlation
functions in Table~\ref{tab:table}. We find that, in spite of the
presence of the trap, the correlation functions still show exponential
or algebraic scaling away from the edge of the lattice. In fact, a
correlation function can have different decay behavior in different
parts of the trap. We also find that SF, PSF, and CFSF still exist.
The remainder of this article focusses on experimental signatures
that distinguish between these orders by calculating the density distibution,
the time-of-flight image after an expansion, or the structure factor
for Bragg spectroscopy.

\begin{figure}
\includegraphics[width=4.43cm,height=4.5cm]{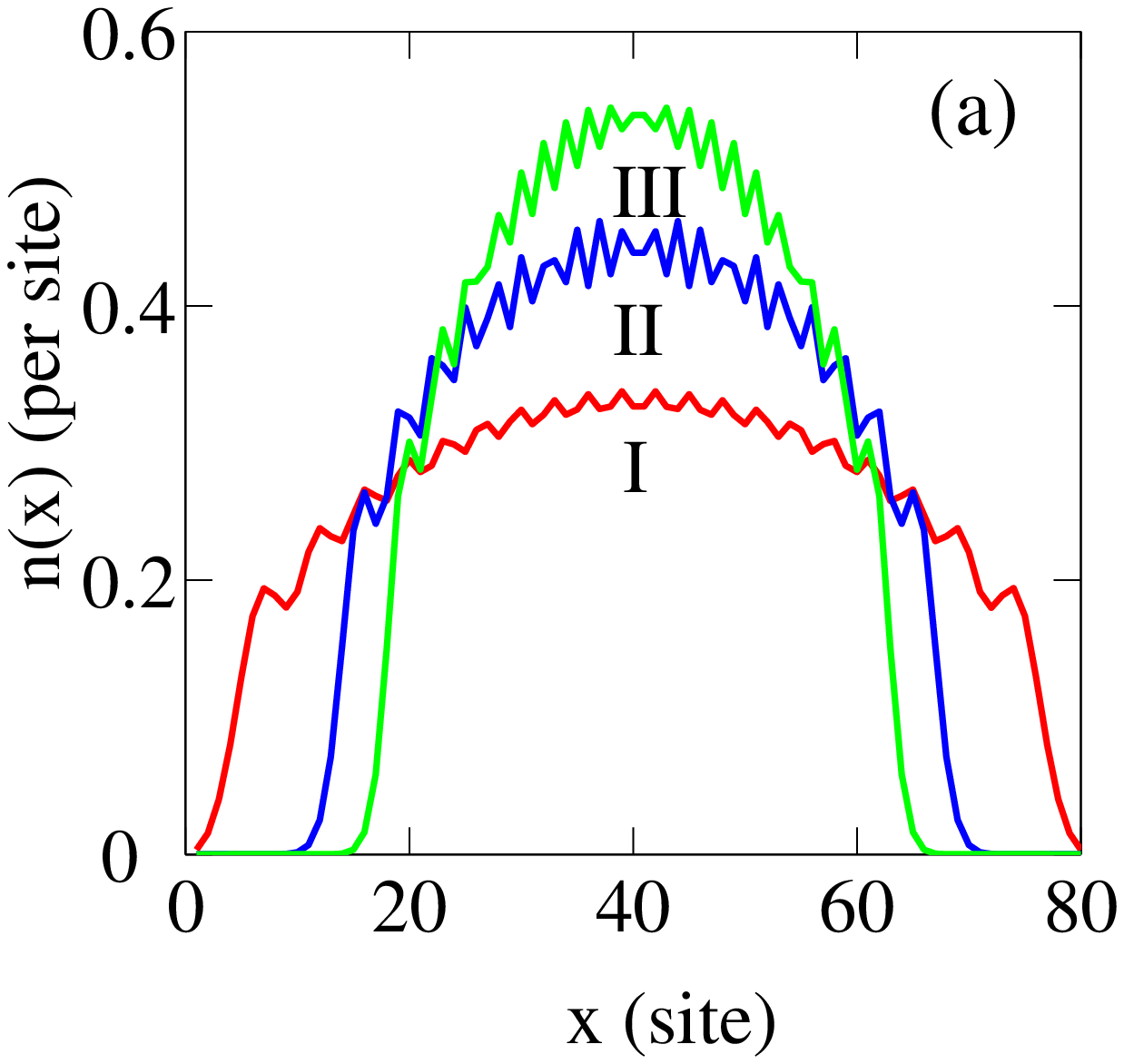}\includegraphics[width=4.5cm,height=4.4cm]{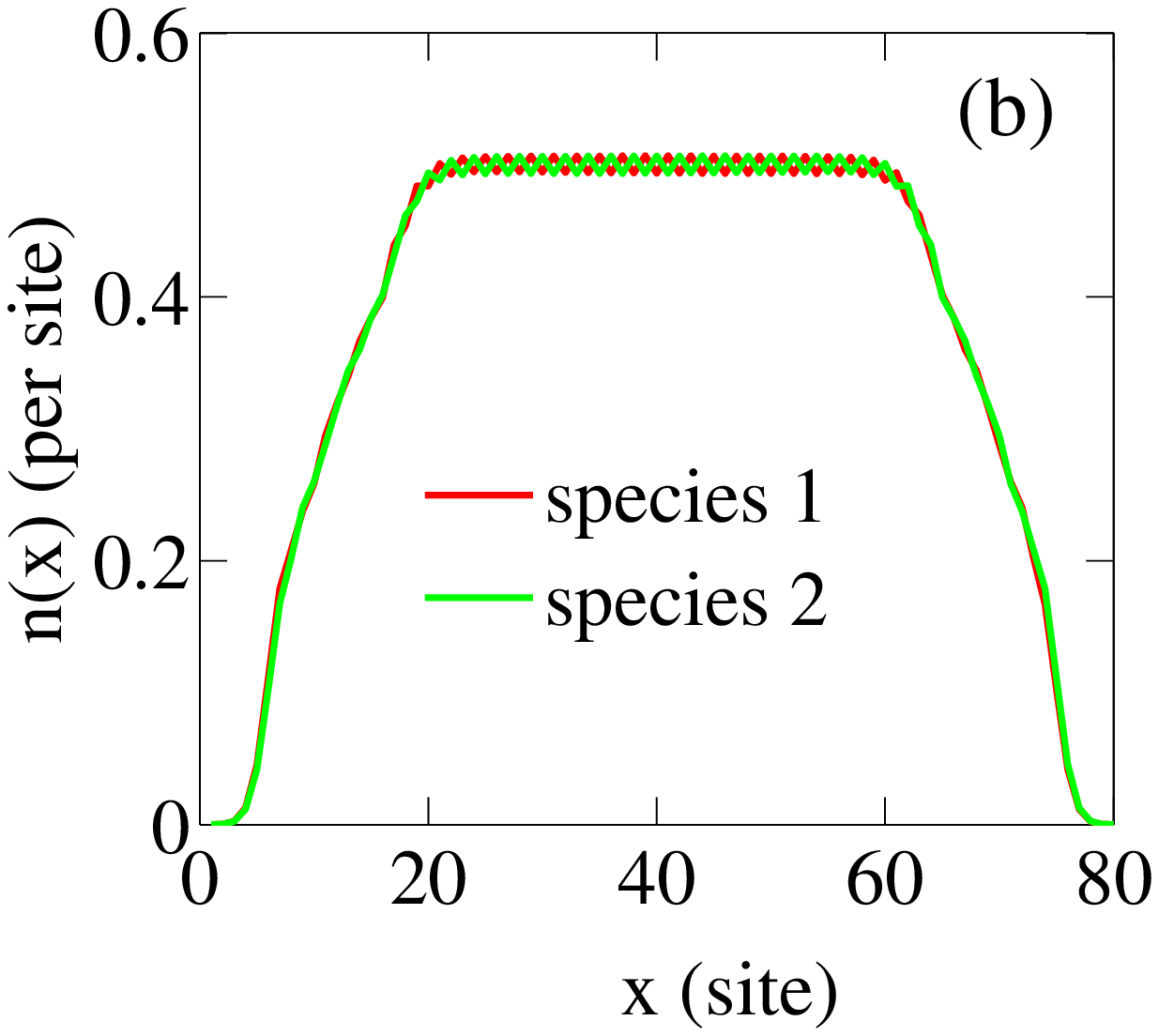}

\caption{\label{fig:density of CFSF,PSF&SF}Density distribution of a trapped
system for $t=0.02U$. (a) Attractive interaction $U_{12}$. The trap
frequency is $\Omega=1\times10^{-5}U$ and the number of atoms is
$20$ for each species. For attractive interactions, the density distributions
of the two species are identical. For $U_{12}=-0.01U$ (curve I) the
system is superfluid. For $U_{12}=-0.11U$ (curve II) and $U_{12}=-0.21$U
(curve III), the system is in the paired superfluid (PSF) state. As
$U_{12}$ becomes more negative the distribution gradually shrinks
in size. (b) Repulsive interaction $U_{12}=0.2U$ with $\Omega=8\times10^{-5}U$
and $30$ atoms of each species. The red and green curves correspond
to the species, respectively. The density distribution has a 'plateau'
with half-filling in the center of the trap. The system is in a counter-flow
superfluid (CFSF) state. The two species have weak interlocked density
modulations around half filling. }

\end{figure}

\textit{Density distribution:} We find that in a trapped system PSF
and CFSF can only exist when the density distribution satisfies certain
conditions. For PSF, the density of each species at the center of
the trap, $n_{{\rm center}}$, must be less than one atom per site
or equivalently per lattice constant $a_{L}$. (The density is largest
at the center.) For CFSF, $n_{{\rm center}}$ must satisfy $n_{{\rm center}}a_{L}=1/2$.
Once such conditions are satisfied, the critical value of $U_{12}$
for PSF and CFSF is close to the one for a homogeneous system (See
Figs.~\ref{fig:phasedia} and \ref{fig:The-phase-diagram at half filling}).

In Fig.~\ref{fig:density of CFSF,PSF&SF}(a) we show density distributions
for three attractive interactions $U_{12}$ and a hopping parameter
equal to the one used for Fig.~\ref{fig:phasedia}. For all attractive
interactions, the density distributions of each species are the same.
For more attractive inter-species interaction, the density distribution
concentrates near the center of the trap. There is no discontinuous
change in the density distribution when the system goes from SF to
PSF.

In Fig.~\ref{fig:density of CFSF,PSF&SF}(b) we show the density
distribution for $U_{12}=0.2U$. In this case in the center of the
trap, where the density distribution is constant or has a {}``plateau'',
the system is in a CFSF state. The {}``plateau'' is at half-filling
consistent with predictions from a local density approximation and
noting that in Fig.~\ref{fig:phasedia} CFSF only occurs at $\nu=1/2$.
Towards the edge, where the density is decreasing sharply, it is in
a SF state. The plateau implies that the system is incompressible
in the center.


%
\begin{figure}
\includegraphics[width=4.5cm,height=4.4cm]{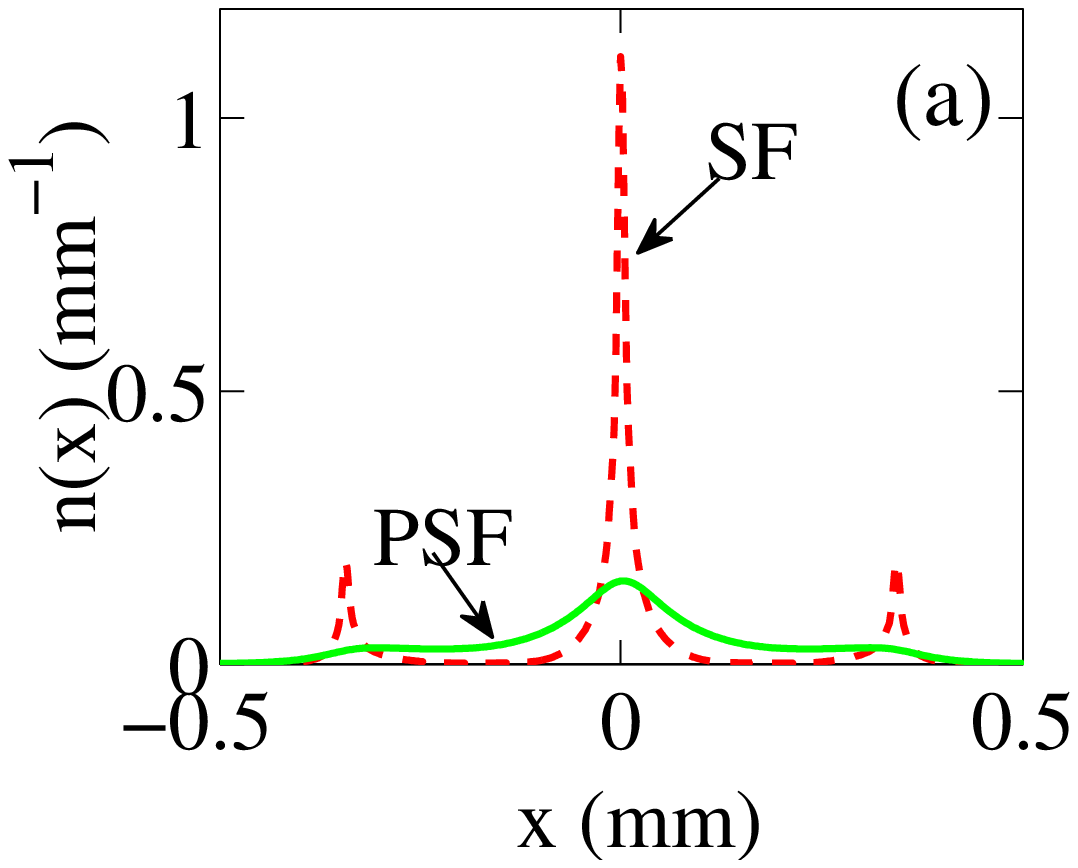}\includegraphics[width=4.5cm,height=4.4cm]{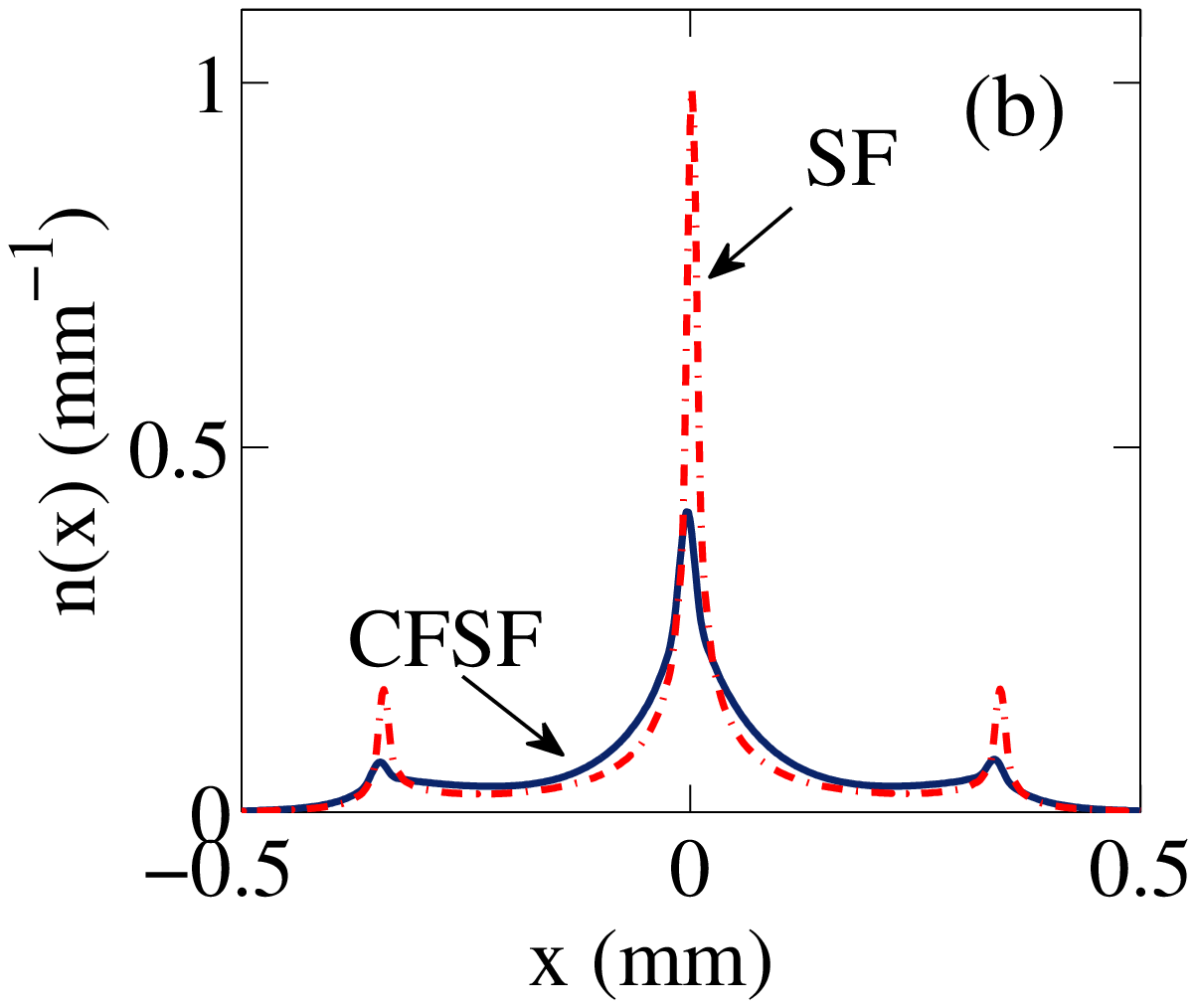}

\caption{\label{fig:TOF_PSF/CFSF/SF} Density distribution after a time-of-flight
expansion. We assume $^{87}$Rb atoms and use an expansion time of
0.03s. The hopping energy is $t=0.02U$. Panel (a): For attractive
interaction $U_{12}$, we show the TOF expansion of a SF state at
$U_{12}=-0.01U$ (red line) and of a PSF state at $U_{12}=-0.21U$
(green line). The two curves correspond to the expansion of the densities
shown as curve I and III in Fig.~\ref{fig:density of CFSF,PSF&SF}(a)
The trap frequency is $\Omega=1\times10^{-5}U$. Panel (b): For repulsive
interaction, we show a TOF expansion of a SF state at $U_{12}=0.01U$
and of a CFSF state at $U_{12}=0.21U$. The trap frequency is $\Omega=8\times10^{-5}U$.}

\end{figure}

\textit{Time of flight measurement:} A widely used measurement technique
in the field of ultra-cold atoms is measuring the density of atoms
after a time-of-flight (TOF) expansion. The 1D optical lattice potential
and the harmonic trap are abruptly turned off at time $T=0$ and the
atoms expand freely afterwards. We calculate the density at time $T$,
according to \begin{equation}
n_{a}(x,T)=\langle c_{a}^{\dagger}(x,T)c_{a}(x,T)\rangle\end{equation}
 with $a=1,2$. The operators $c_{a}(x,T)$ are related to the lattice
operator $b_{a,j}$ according to \begin{equation}
b_{a}(x,T)=\sum_{j=1}^{N}w(x-r_{j},T)b_{a,j},\end{equation}
 where $w(x,T)=\sqrt{d/\sqrt{2\pi}\Delta(T)^{2}}\exp(-x^{2}/(4\Delta(T)^{2}))$
describes the free expansion from the initial Gaussian wavefunction
of an atom in a lattice site and $\Delta(T)^{2}=d^{2}+iT\hbar/(2m)$.
The parameter $d$ is the width of the initial Gaussian state and
$m$ is the atomic mass. The density distribution $n_{a}(x,T)$ is
then given by 
\[
n_{a}(x,T)=\sum_{j_{1},j_{2}=1}^{N}w^{*}(x-r_{j_{1}},T)w(x-r_{j_{2}},T)G(j_{1},j_{2}),\]
where $G(j_{1},j_{2})$ is the single-particle Green's function. In
Fig.~\ref{fig:TOF_PSF/CFSF/SF} we show examples of TOF expansions
of PSF, CFSF, and SF order. For the SF phase, we find a strongly peaked
interference pattern, reflecting the single-particle quasi-long range
order. For both PSF and CFSF phases, the TOF density shows a broad
Lorentzian distribution, which is due to the exponential decay of
the single-particle Green's function.

\textit{Feshbach ramp:} In order to detect the superfluidity of pairs,
we consider applying a Feshbach ramp to pairwise project the atoms
onto molecules formed by one atom from each species, which is similar
to detection of fermionic pairs in the BCS regime~\cite{BCS-BEC}.
In those experiments, a fast ramp across a Feshbach resonance was
used, followed by a time-of-flight expansion. The density distribution
of the molecules showed the superfluidity of fermionic pairs. We propose
a similar detection for bosonic pairs in PSF.

\begin{figure}[b]
 \includegraphics[width=8cm]{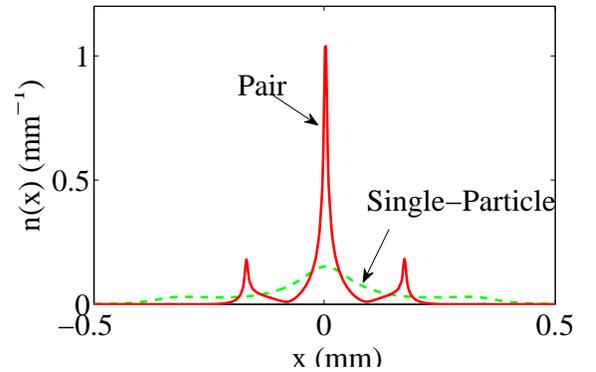}

\caption{\label{fig:TOF_molecules}Density distribution of molecules after
time-of-flight expansion of state III in Fig.~\ref{fig:density of CFSF,PSF&SF}(a).
The expansion time is 0.03s. We assume two hyperfine states of $^{87}Rb$.
These are converted into Feshbach molecules at $T=0$ via a fast ramp
across a resonance. We assume a complete conversion. The strongly
peaked interference pattern of molecules indicates the presence of
a quasi-condensate of pairs. For comparison, we also show the TOF
expansion of atoms in the PSF phase for the same parameters. The broad
Lorentzian distribution demonstrates the absence of single-particle
SF. }

\end{figure}

To give a simple estimate of a TOF image after a Feshbach ramp, we
imagine that bosons of different species on the same lattice site
are converted into molecules. This leads to the replacement $b_{1,j}b_{2,j}\rightarrow M_{j}$,
where $M_{j}$ is the molecule annihilation operator. A TOF density
of the molecules at position $x$ and time $T$ is given by \begin{equation}
n_{M}(x,T)=\sum_{j_{1},j_{2}=1}^{N}w^{*}(x-r_{j_{1}},T)w(x-r_{j_{2}},T)R_{s}(j_{1},j_{2}).\label{eq:M_n}\end{equation}
 In the expanding wave function $w(x,T)$, the mass $m$ is replaced
by the mass of the molecule. We assume the same initial width $d$.
In a more realistic estimate, the conversion efficiency to molecules
would not be 100\%, but approximately given by the square of the overlap
of the molecular wave function and the single-atom wave functions.
This leads to a reduced signal. The spatial dependence, however, remains
the same. In Fig.~\ref{fig:TOF_molecules}, we see an example of
the density of molecules after TOF and, for comparison, the atomic
density after TOF for the PSF state. The strongly peaked molecular
distribution indicates the quasi-condensate of the bosonic pairs.
The single-atom density is a broad Lorentzian distribution, indicating
the absence of single-particle SF.

\begin{figure}
\includegraphics[width=7.5cm]{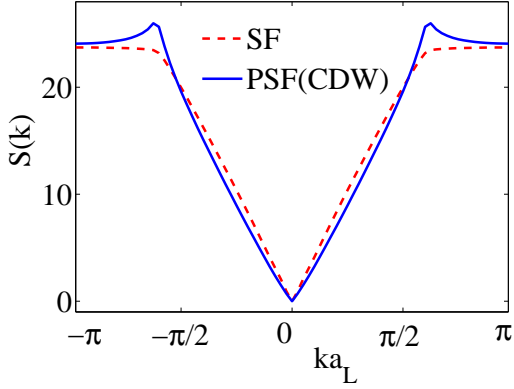}

\caption{\label{fig:sk}Structure factor at filling $\nu=0.3$. For $U_{12}=-0.01U$
the system is in the SF regime (dashed line) and for $U_{12}=-0.07U$
the system is in the PSF regime (continuous line). Cusps at $|k|=2\pi\nu$
only occur for $U_{12}=-0.07U$ indicating the coexistence of CDW
with PSF order.}

\end{figure}

\textit{Bragg spectroscopy:} To detect the presence of CDW order,
one can use Brag spectroscopy \cite{StengerStruc,Bragg}. The quantity
that is measured in those experiments is either the dynamic or static
structure factor. Here we calculate the static structure factor $S_{a}(k)$
for species $a=1,2$. It is defined as \begin{eqnarray}
S_{a}(k) & = & \frac{1}{N}\sum_{j_{1},j_{2}}e^{-ika_{L}(j_{1}-j_{2})}(\langle n_{a}(j_{1})n_{a}(j_{2})\rangle\nonumber \\
 &  & \quad\quad\quad\quad-\langle n_{a}(j_{1})\rangle\langle n_{a}(j_{2})\rangle)\,.\label{eq:Sk}\end{eqnarray}
 For wavevectors $k$ near twice the {}``Fermi wavevector'' $k_{F}$,
the structure factor $S(k)\sim||k|-2k_{F}|^{1-\alpha_{CDW}}$ with
$\alpha_{CDW}=2-K_{S}-K_{A}$ \cite{giamarchi_book}. In our system,
$K_{S}+K_{A}$ is always larger than 1 and, thus, $1-\alpha_{CDW}$
is positive. Consequently, the structure factor does not diverge.
In the CDW regime with $K_{S}+K_{A}<2$ the power $1-\alpha_{CDW}$,
however, is less than one. This gives $S(k)$ cusps at $\pm2k_{F}$
when CDW quasi-long range order is present. In Fig.~\ref{fig:sk}
we show examples of $S(k)$ for a case with and without CDW.

\begin{figure}[b]
 \includegraphics[width=7.5cm]{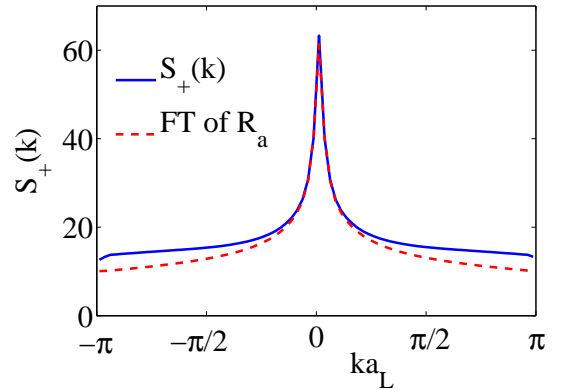}

\caption{\label{fig:pi/2}Structure factor $S_{+}(k)$ (blue line) after applying
a $\pi/2$ pulse in the CFSF phase. The quasi-condensate of anti-pairs
generates an algebraic peak at $k=0$. The cusp also appear in the
Fourier transform of the anti-pair correlation function $R_{a}(i,j)=\langle b_{1,i}^{\dagger}b_{2,i}b_{2,j}^{\dagger}b_{1,j}\rangle$(red
dashed line). }

\end{figure}

\textit{Bragg Spectroscopy preceded by a $\pi$/2 pulse:} To detect
CFSF order, we propose the following detection method. It applies
to the case that the mixture is composed of atoms in different internal
states rather than different atomic species. First, we apply a $\pi/2$
pulse, which transfers the atoms into the superpositions $b{}_{1/2,i}\rightarrow b_{\pm,i}=(b_{1,i}\pm b_{2,i})/\sqrt{2}$.
We then measure the structure factor, which now corresponds to the
Fourier transform of the density correlations $R_{n\pm}(i,j)=\langle n{}_{\pm,i}n{}_{\pm,j}\rangle-\langle n_{\pm,i}\rangle\langle n_{\pm,j}\rangle$.
In terms of the original $b_{1/2,i}$ operators these density correlations
are given by \begin{eqnarray}
R_{n\pm}(i,j) & = & \frac{1}{4}\langle(n_{1,i}+n_{2,i})(n_{1,j}+n_{2,j})\rangle\nonumber \\
 &  & \quad-\frac{1}{4}(\langle n_{1,i}\rangle+\langle n_{2,i}\rangle)(\langle n_{1,j}\rangle+\langle n_{2,j}\rangle)\nonumber \\
 &  & \quad\quad+\frac{1}{2}\langle b_{1,i}^{\dagger}b_{2,i}b_{2,j}^{\dagger}b_{1,j}\rangle\label{eq:nn}\end{eqnarray}
 The last term in the above equation is the correlation function $R_{a}(i,j)$
of the order parameter of CFSF, $b_{1,j}b_{2,j}^{\dagger}$. In Fig.~\ref{fig:pi/2},
we show the structure factor $S_{+}(k)$, the Fourier transform of
Eq.~\ref{eq:nn}, as well as the Fourier transform of $R_{a}(i,j)$.
Both $S_{+}(k)$ and the Fourier transform of $R_{a}(i,j)$ have a
cusp around $k=0$. The cusp is due to the long-range correlations
of the anti-pairs in the CFSF. The two functions are nearly identical
near $k=0$, indicating that the momentum distibution of anti-pairs
can be measured by determining the structure factor $S_{+}(k)$ .

\section{\label{sec:Summary}Summary}

We have studied ground state properties of one-dimensional Bose mixtures
in an optical lattice using both Tomonaga-Luttinger liquid theory
and the time-evolving block decimation method. We first discussed
the zero-temperature phase diagram in a homogeneous system at different
filling fractions and different parameter regimes. We have shown that
1D Bose mixtures in an optical lattice can have quasi-long range orders
that include superfluid, paired superfluid (PSF), counterflow superfluid
(CFSF), and Mott insulator. We also found that each type of superfluid
order can coexist with charge density wave (CDW) order and that in
both PSF and CFSF phases single particle superfluidity (SF) is absent.

In addition, we discussed ways of realizing and detecting these phases
experimentally. We propose using a Feshbach ramp to probe the momentum
distribution of pairs in the PSF, which shows signatures of the quasi-condensate
of pairs. To detect the CFSF for a mixture composed of two atomic
hyperfine states, we propose to measure the static structure factor
by using Bragg spectroscopy preceded by a $\pi/2$ pulse. A sharp
peak in the structure factor was shown to be dominated by the contribution
from the momentum distribution of anti-pairs in the CFSF phase. Finally,
we suggest to detect CDW order with Bragg spectroscopy.

This work was supported by NSF under Physics Frontier Grant PHY-0822671.
L.M. acknowledges support from an NRC/NIST fellowship. I.D. acknowledges
support from a Grant-in-Aid from JSPS.

\appendix

\section{\label{sec:The-effective-one-species} TEBD method for two-species
many-body systems}

In this appendix, we briefly review the time-evolving block decimation
(TEBD) method~\cite{Vidal} used in Sec.~\ref{sec:Numerical-Approach}
and explain an efficient way to apply the TEBD to a two-species Bose-Hubbard
model. We use the number-conserving version of the TEBD method~\cite{daley}.

The TEBD determines the ground state via an imaginary time evolution
for one-dimensional (1D) quantum lattice systems. In this method the
Hilbert space $\mathbf{H}$ is decomposed as \begin{equation}
\mathbf{H}=\otimes_{l=1}^{M}\mathbf{H}_{l}.\end{equation}
 Here, $l$ refers to the $l$th lattice site, $M$ is the number
of sites, and $\mathbf{H}_{l}$ is the local Hilbert space at site
$l$ with local dimension $d$, independent of $l$. Any state $|\Psi\rangle$
in $\mathbf{H}$ is represented as \begin{eqnarray}
|\Psi\rangle=\sum_{j_{1},j_{2},\ldots,j_{M}=1}^{d}c_{j_{1},j_{2},\ldots,j_{M}}|j_{1}\rangle|j_{2}\rangle\cdots|j_{M}\rangle.\end{eqnarray}
 In the TEBD algorithm, coefficients $c_{j_{1},j_{2},\ldots,j_{M}}$
are decomposed as \begin{eqnarray}
c_{j_{1},j_{2},\ldots,j_{M}}\!\!\! & = & \!\!\!\sum_{\alpha_{1}=1}^{\chi_{1}}\sum_{\alpha_{2}=1}^{\chi_{2}}\cdots\sum_{\alpha_{M-1}=1}^{\chi_{M-1}}\Gamma_{\alpha_{1}}^{[1]j_{1}}\lambda_{\alpha_{1}}^{[1]}\Gamma_{\alpha_{1}\alpha_{2}}^{[2]j_{2}}\lambda_{\alpha_{2}}^{[2]}\cdots\nonumber \\
 &  & \times\lambda_{\alpha_{M-2}}^{[M-2]}\Gamma_{\alpha_{M-2}\alpha_{M-1}}^{[M-1]j_{M-1}}\lambda_{\alpha_{M-1}}^{[M-1]}\Gamma_{\alpha_{M-1}}^{[M]j_{M}}.\label{eq:tensorproduct}\end{eqnarray}
The variables $\lambda_{\alpha_{l}}^{[l]}$ and $\chi_{l}$ are the
Schmidt coefficients and rank of the Schmidt decomposition of $|\Psi\rangle$
with respect to the bipartite splitting of the system into $[1,\ldots,l-1,l]:[l+1,l+2,\ldots,M]$,
\begin{eqnarray}
|\Psi\rangle=\sum_{\alpha_{l}=1}^{\chi_{l}}\lambda_{\alpha_{l}}^{[l]}|\Phi_{\alpha_{l}}^{[1,\ldots,l-1,l]}\rangle|\Phi_{\alpha_{l}}^{[l+1,l+2,\ldots,M]}\rangle.\end{eqnarray}
 We take $\lambda_{\alpha}^{[l]}>\lambda_{\beta}^{[l]}$ for all $\alpha<\beta$.
In one dimension, the rank $\chi_{l}$ at the center of the system
must be of the order $d^{M/2}$ in order to express arbitrary states.
However, since it is empirically known that the Schmidt coefficients
$\lambda_{\alpha}^{[l]}$ decrease rapidly with index $\alpha$ for
the ground and low-lying excited states, we set $\chi_{l}$ to a relatively
small number $\chi$ for all $l$.

To efficiently simulate the two-species Bose-Hubbard model (Eq.~\ref{eq:hamiltonian}
in the main text), we map it onto the one-species Hamiltonian \begin{eqnarray}
H & = & -t\sum_{l=1}^{2N-2}(b_{l}^{\dagger}b_{l+2}+{\rm h.c.})+U_{12}\sum_{{\rm odd}\, l}n_{l}n_{l+1}\nonumber \\
 &  & +\frac{U}{2}\sum_{l=1}^{2N}n_{l}(n_{l}-1),\label{eq:1speBHH}\end{eqnarray}
 where $N$ is the number of sites in the original two-species Hamiltonian.
In this one-species Hamiltonian, there are $2N$ sites, each of which
is indexed by $l$. The odd sites $l$ correspond to species 1 and
the even sites to species 2. Hopping between neighboring sites $-t\, b_{a,i}^{\dagger}b_{a,i+1}$
in Eq.~\ref{eq:hamiltonian} is mapped onto a next-nearest-neighbor
hopping $-t\, b_{l}^{\dagger}b_{l+2}$ in Eq.~\ref{eq:1speBHH}.
Similarly, the inter-species onsite-interaction $U_{12}n_{1,i}n_{2,i}$
is mapped onto the nearest-neighbor interaction $U_{12}n_{l}n_{l+1}$.
This type of mapping has been successfully applied to treat the two-legged
Bose-Hubbard model~\cite{danshita1}.

We map the two-species Bose-Hubbard Hamiltonian Eq.~\ref{eq:hamiltonian}
onto the one-species Hamiltonian because it reduces computational
cost dramatically. This cost in TEBD~\cite{Vidal} scales as $Md^{3}\chi^{3}$.
For the two-species system with $N$ sites we must define a dimension
of the local Hilbert space for each species, say $D$. Hence, at each
site there are $D^{2}$ basis functions and the cost scales as $ND^{6}$.
On the other hand, for the mapped Hamiltonian with $2N$ sites and
a local dimension $D$ the cost only scales as $2ND^{3}$. In our
calculation, we set $d=3$ for the filling factor $\nu\leq0.8$ and
$d=5$ for $\nu=0.9,1$. In this case, the mapping makes the computation
five to ten times faster.

Imaginary time evolution of any state to the ground state is given
by repeated application of $e^{-iH\delta}$ on $|\Psi\rangle$, where
$\delta$ is a small imaginary time step. To apply this operator we
first split the Hamiltonian into three parts as $H=H_{{\rm int}}+H_{{\rm hop}}^{{\rm odd}}+H_{{\rm hop}}^{{\rm even}}$,
where

\begin{eqnarray}
H_{{\rm int}} & = & \sum_{m=1}^{N}\left[U_{12}n_{2m-1}n_{2m}+Un_{2m-1}(n_{2m-1}-1)\right.\nonumber \\
 &  & \left.+Un_{2m}(n_{2m}-1)\right],\\
H_{{\rm hop}}^{{\rm odd}} & = & -t\sum_{{\rm odd}\, m}(b_{2m-1}^{\dagger}b_{2m+1}+b_{2m}^{\dagger}b_{2m+2}+{\rm h.c.}),\nonumber \\
H_{{\rm hop}}^{{\rm even}} & = & -t\sum_{{\rm even\, m}}(b_{2m-1}^{\dagger}b_{2m+1}+b_{2m}^{\dagger}b_{2m+2}+{\rm h.c.}).\nonumber \end{eqnarray}
Subsequently, we use the second-order Suzuki-Trotter expansion to
decompose $e^{-i\hat{H}\delta}$ as 
\begin{eqnarray}
e^{-iH\delta} & = & e^{-iH_{{\rm int}}\delta/2}e^{-iH_{{\rm hop}}^{{\rm odd}}\delta/2}e^{-iH_{{\rm hop}}^{{\rm even}}\delta}e^{-iH_{{\rm hop}}^{{\rm odd}}\delta/2}\nonumber \\
 &  & \times e^{-iH_{{\rm int}}\delta/2}+O(\delta^{3}),\end{eqnarray}
Each of the operators $e^{-iH_{{\rm int}}\delta/2}$, $e^{-iH_{{\rm hop}}^{{\rm odd}}\delta/2}$,
and $e^{-i\hat{H}_{{\rm hop}}^{{\rm even}}\delta}$ can be decomposed
into a product of two-site operators, which can be efficiently applied
to the matrix product state $|\Psi\rangle$~\cite{Vidal,shi,danshita2}.
We use swapping techniques to apply the next-nearest-neighbor operators
$e^{-iH_{{\rm hop}}^{{\rm odd}}\delta/2}$ and $e^{-i\hat{H}_{{\rm hop}}^{{\rm even}}\delta}$
~\cite{shi,danshita2}. 

\begin{thebibliography}{10}
\bibitem{BEC}S. Bose, \textit{Z. Phys. 26}, \textbf{178} (1924);
A. Einstein, \textit{Sitzungsber. K. Preuss. Akad. Wiss., Phys. Math.
Kl.}, \textbf{261} (1924); \textit{Sitzungsber. K. Preuss. Akad. Wiss.,
Phys. Math. Kl.}, \textbf{3} (1925).

\bibitem{BEC-exp}K. B. Davis, M. -O. Mewes, M. R. Andrews, N. J.
van Druten, D. S. Durfee, D. M. Kurn and W. Ketterle, \textit{Phys.
Rev. Lett.} \textbf{75} 3969 (1995); M.H. Anderson, J. R. Ensher,
M. R. Matthews, C. E. Wieman and E. A. Cornell, \textit{Science} \textbf{269}
198 (1995); C. C. Bradley, C. A. Sackett, J. J. Tollett, R. G. Hulet,
\textit{Phys. Rev. Lett.} \textbf{75} 1687 (1995).

\bibitem{BCS}L.N. Cooper, \textit{Phys. Rev.} \textbf{104} 1189 (1956);
J. Bardeen, L. N. Cooper and J. R. Shrieffer, \textit{Phys. Rev.}
\textbf{106} 162 (1957); \textit{Phys. Rev.} \textbf{108} 1175 (1957).

\bibitem{BCS-BEC}C. A. Regal, M. Greiner and D. S. Jin, \textit{Phys.
Rev. Lett.} \textbf{92} 040403 (2004); M.W. Zwierlein, C. A. Stan,
C. H. Schunck, S. M. F. Raupach, A. J. Kerman and W. Ketterle, \textit{Phys.
Rev. Lett.} \textbf{92} 120403 (2004); M. Bartenstein, A. Altmeyer,
S. Riedl, S. Jochim, C. Chin, J. Hecker Denschlag and R. Grimm, \textit{Phys.
Rev. Lett.} \textbf{92} 120401 (2004); T. Bourdel, L. Khaykovich,
J. Cubizolles, J. Zhang, F. Chevy, M. Teichmann, L. Tarruell, S. J.
J. M. F. Kokkelmans and C. Salomon \textit{Phys. Rev. Lett.} \textbf{93}
050401 (2004).

\bibitem{KuklovCFSF}A. B. Kuklov and B. V. Svistunov, \textit{Phys.
Rev. Lett.} \textbf{90}, 100401 (2003).

\bibitem{KuklovDiagram}A. Kuklov, N. Prokof'ev and B. Svistunov,
\textit{Phys. Rev. Lett.} \textbf{92}, 050402 (2004);\textit{Phys.
Rev. Lett.} \textbf{92}, 030403 (2003).

\bibitem{Bose-Hubbard}D. Jaksch, \textit{Phys. Rev. Lett.} \textbf{81}
, 3108 (1998)

\bibitem{MI-SF(3D)}M. Greiner, O. Mandel, T. Esslinger, T. W. Hänsch
and I. Bloch, \emph{Nature} \textbf{415}, 39 (2002)\emph{.}

\bibitem{MI-SF(1D)}T. Stöferle, H. Moritz, M. Köhl and T. Esslinger,
\textit{\emph{Phys. Rev. Lett.}} \textbf{\emph{92}},130403(2004)

\bibitem{MI-SF(2D)}I. B. Spielman, W. D. Phillips, and J. V. Porto
, \emph{Phys. Rev. Lett.} \textbf{98}, 080404 (2007)

\bibitem{BlochTonks}B. Paredes, A. Widera, V. Murg, O. Mandel, S.
Foelling, I. Cirac, G. Shlyapnikov, T. W. Hansch and I. Bloch, \textit{Nature}
\textbf{429}, 277 (2004).

\bibitem{BlochSpin}A. Widera, S. Trotzky, P. Cheinet, S. Fölling,
F. Gerbier, and I. Bloch , \textit{Phys. Rev. Lett.} \textbf{100},
140401 (2008)

\bibitem{giamarchi_book} T. Giamarchi, \textit{Quantum Physics in
one dimension}, (Oxford Univ. Press, Oxford, UK, 2004)

\bibitem{Vidal}G. Vidal, \textit{Phys. Rev. Lett} \textbf{98}, 070201
(2007); G. Vidal, \textit{ibid.} \textbf{91}, 147902 (2003); \emph{ibid.}
\textbf{93}, 040502 (2004); S.R. White and A. E. Feiguin, \textit{ibid.}
\textbf{93}, 076401 (2004);

\bibitem{shi} Y.-Y. Shi, L.-M. Duan, and G. Vidal, \emph{Phys. Rev.
A} \textbf{74}, 022320 (2006).

\bibitem{daley} A. J. Daley, S. R. Clark, D. Jaksch, and P. Zoller,
\emph{Phys. Rev. A} \textbf{72}, 043618 (2005); A. J. Daley, Ph.D.
thesis, Leopold-Franzens-Universität Innsbruck, 2005 (unpublished).

\bibitem{danshita1}I. Danshita, J. E. Williams, C. A. R. Sá de Melo,
and C. W. Clark, \emph{Phys. Rev. A} \textbf{76}, 043606 (2007); I.
Danshita, C. A. R. Sá de Melo, and C. W. Clark, \emph{Phys. Rev. A}
\textbf{77}, 063609 (2008).

\bibitem{danshita2} I. Danshita and P. Naidon, \emph{Phys. Rev. A}
\textbf{79}, 043601 (2009).

\bibitem{Santos}A. Argüelles and L. Santos, Phys. Rev. A 75, 053613
(2007); ibid 77, 059904 (2008).

\bibitem{supersolid_LM}L. Mathey, I. Danshita and C. W. Clark, \textit{Phys.
Rev. A} \textbf{79}, 011602(R) (2009)

\bibitem{supersolid}G. G. Batrouni, F. Hébert and R. T. Scalettar,
\textit{Phys. Rev. Lett.} \textbf{97}, 087209 (2006); V. W. Scarola
and S. Das Sarma, \emph{ibid.} \textbf{\emph{95}}, 033003 (2005);
P. Sengupta1, L. P. Pryadko1, F. Alet, M. Troyer and Guido Schmid,
\emph{ibid.} \textbf{94}, 207202 (2005); S. Wessel and M. Troyer,
\emph{ibid.} \textbf{95}, 127205 (2005); D. Heidarian and K. Damle,
\emph{ibid} \textbf{95}, 127206 (2005); R. G. Melko, A. Paramekanti,
A. A. Burkov, A. Vishwanath, D. N. Sheng, and L. Balents, \emph{ibid.}
\textbf{95}, 127207 (2005); H.P. Büchler and G. Blatter, \emph{ibid.}
\textbf{91}, 130404 (2004); M. Boninsegni and N. Prokof'ev, \emph{ibid.}
\textbf{95}, 237204 (2005); M. Boninsegni, \textit{J. Low. Temp. Phys.}
\textbf{132}, 39 (2005); D. L. Kovrizhin, G. Venketeswara Pai and
S. Sinha, \textit{Euro. Phys. Lett.} \textbf{72}, 162 (2005); F. Karim
Pour, M. Rigol, S. Wessel, and A. Muramatsu, \textit{Phys. Rev. B}
\textbf{75}, 161104 (2007).

\bibitem{Commensurate}L. Mathey, \textit{Phys. Rev. B} \textbf{75},
144510 (2007).

\bibitem{Pair-SS-Lewenstein}C. Trefzger, C. Menotti, M. Lewenstein,
arXiv:0904.1552.

\bibitem{Mori}A. Masaki, S. Tsukada, and H. Mori, \emph{J. Phys.
Conf. Ser.} \textbf{150}, 032050 (2009).

\bibitem{Haldane} F. D. M. Haldane, \textit{Phys. Rev. Lett.} \textbf{47},
1840 (1981).

\bibitem{Cazalilla} M.A. Cazalilla, \textit{J. Phys. B: At. Mol.
Opt. Phys.} \textbf{37}, S1 (2004).

\bibitem{CazalillaTonks} M.A. Cazalilla, \textit{Phys. Rev. A} \textbf{70},
041604(R) (2004).

\bibitem{SG}J. B. Kogut, \textit{Rev. Mod. Phys.} \textbf{51}, 659
(1979).

\bibitem{BKT}J. M. Kosterlitz and D. J. Thouless, \emph{J. Phys.
C} \textbf{6}, 1181 (1973); V. L. Berezinskii, Zh. Eksp. Teor. Fiz.
\textbf{6}, 907 (1970) {[}Sov. Phys. JETP \textbf{32}, 493 (1971)].

\bibitem{Mishra}T. Mishra, R. V. Pai and B. P. Das, \textit{Phys.
Rev. A} \textbf{76}, 013604 (2007).

\bibitem{StengerStruc}J. Stenger, S. Inouye, A. P. Chikkatur, D.
M. Stamper-Kurn, D. E. Pritchard, and W. Ketterle, \textit{Phys. Rev.
Lett.} \textbf{\emph{82}}\emph{,} 4569 (1999); J. Steinhauer, R. Ozeri,
N. Katz, and N. Davidson, \emph{ibid.} \textbf{88}, 120407 (2002).

\bibitem{Bragg}D. Clément, N. Fabbri, L. Fallani, C. Fort, and M.
Inguscio, \emph{Phys. Rev. Lett.} \textbf{102}, 155301 (2009)

\bibitem{Cornell}D.S. Hall, M. R. Matthews, C. E. Wieman, and E.
A. Cornell, \textit{Phys. Rev. Lett.} \textbf{81}, 1543 (1998)

\bibitem{Thalhammer}G. Thalhammer, G. Barontini, L. De Sarlo, J.
Catani, F. Minardi, and M. Inguscio, \emph{Phys. Rev. Lett.} \textbf{100},
210402 (2008)

\end{thebibliography}
\end{document}